\documentclass[journal=jacsat,manuscript=article]{achemso}

\usepackage[version=3]{mhchem} 
\usepackage[usenames,dvipsnames]{xcolor}

\usepackage{multirow}
\usepackage{comment}
\usepackage{booktabs}
\usepackage{braket}

\author{Soumi Haldar}
\affiliation[University of Chicago]
{Department of Chemistry, Chicago Center for Theoretical Chemistry, University of Chicago, Chicago, IL 60637, USA.}
\author{Lorenzo A. Mariano}
\author{Alessandro Lunghi}
\email{lunghia@tcd.ie}
\affiliation[TCD]
{School of Physics, CRANN and AMBER Research Centre, Trinity College, Dublin 2, Ireland}
\author{Laura Gagliardi}
\email{lgagliardi@uchicago.edu}
\affiliation[University of Chicago]
{Department of Chemistry, Chicago Center for Theoretical Chemistry, University of Chicago, Chicago, IL 60637, USA.}

\title[An \textsf{achemso} demo]
  {The Role of Electron Correlation Beyond the Active Space in Achieving Quantitative Predictions of Spin-Phonon Relaxation} 
  
\begin{document}

\begin{abstract}
Single-molecule magnets (SMMs) are promising candidates for molecular-scale data storage and processing due to their strong magnetic anisotropy and long spin relaxation times. However, as temperature rises, interactions between electronic states and lattice vibrations accelerate spin relaxation, significantly limiting their practical applications. Recently, ab initio simulations have made it possible to advance our understanding of phonon-induced magnetic relaxation, but significant deviations from experiments have often been observed. The description of molecules' electronic structure has been mostly based on complete active space self-consistent field (CASSCF) calculations, and the impact of electron correlation beyond the active space remains largely unexplored.
In this study, we provide the first systematic investigation of spin-phonon relaxation in SMMs with post-CASSCF multiconfigurational methods, specifically CAS followed by second-order perturbation theory and multiconfiguration pair-density functional theory. Taking Co(II)- and Dy(III)-based SMMs as case studies, we analyze how electron correlation influences spin-phonon relaxation rates across a range of temperatures, comparing theoretical predictions with experimental observations. 
Our findings demonstrate that post-CASSCF treatments make it possible to achieve quantitative predictions for Co(II)-based SMMs. For Dy(III)-based systems, however, accurate predictions require consideration of additional effects, underscoring the urgent necessity of further advancing the study of the effects of electronic correlation in these complex systems.

\end{abstract}

\begin{center}
    \section{TOC Graphic}
\end{center}
\graphicspath{{Figures/}}

\begin{figure}[h!]
\centerline{\includegraphics[width=5in,height=3.7in]{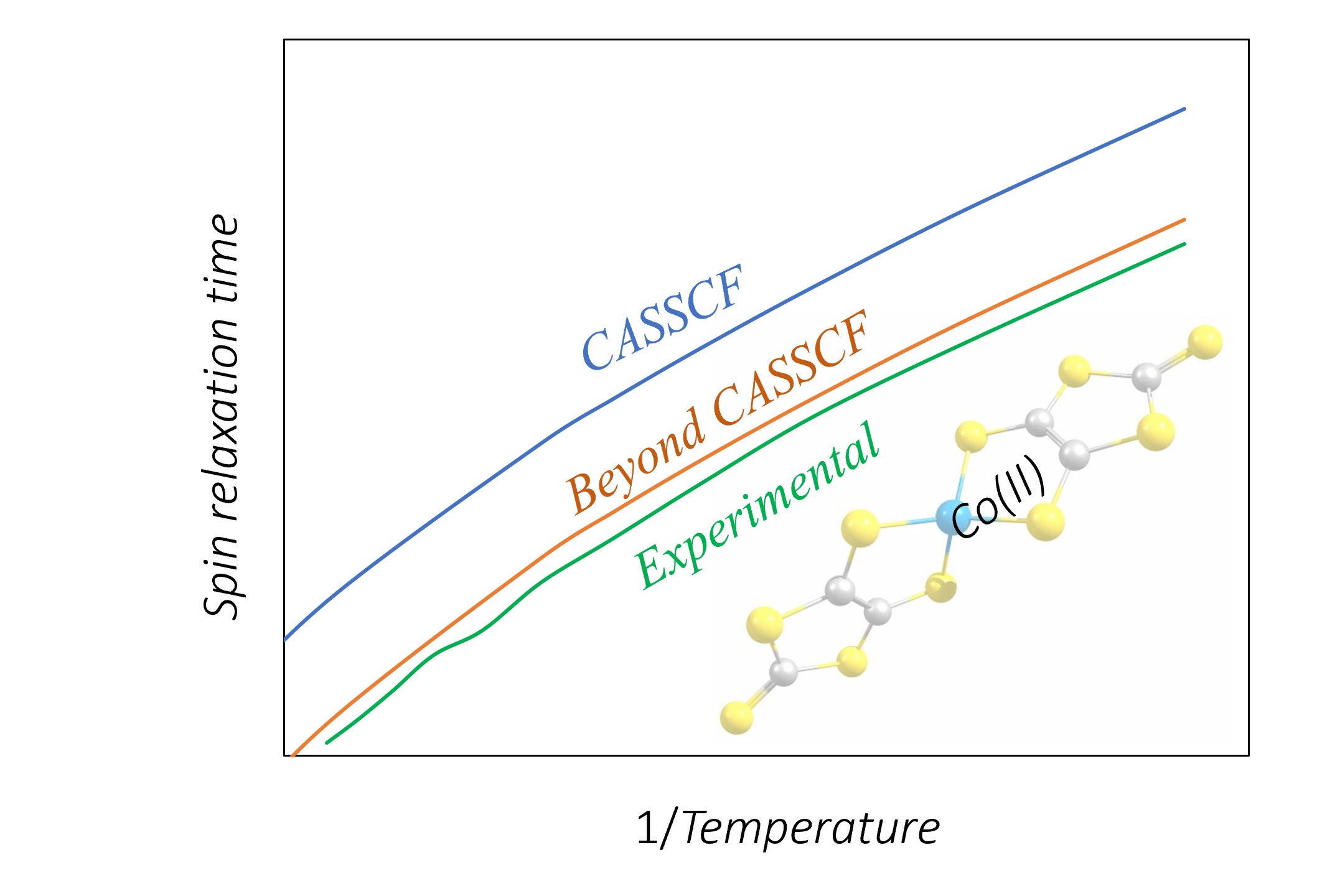}}
\centering
\end{figure}

\section{Introduction}
The inherent magnetic bi-stability exhibited by single-molecule magnets (SMMs) leads to exciting applications of this class of molecules in quantum computing,\cite{leuenberger2001quantum} magnetic data storage,\cite{sessoli1993magnetic} and spintronics\cite{PhysRevB.73.085414}. These systems are distinguished by their ability to retain magnetic information at a molecular level at low temperatures, even after removing the magnetizing field – just like bulk hard ferromagnets. The reason they can retain their magnetization for so long is that the doubly degenerate ground magnetic sublevels with opposite spin orientations are separated by a large energy barrier - because of which the magnetic reversal or spin flip between the two states is very slow. Having a long enough lifetime, the doubly degenerate spin sublevels can be efficiently used as magnetic binary memory units or quantum bits (qubits). However, in the presence of any interactions with their environment, the performance and functionality of SMMs are drastically reduced by a shortened spin lifetime. At finite temperatures, one of the main sources of such perturbation is the interaction between spins and quantized lattice vibrations known as phonons.\cite{D2CS00705C, Lunghi2023} Due to such interaction the electronic or nuclear spins can absorb/emit one or multiple phonons from/into the lattice, eventually bringing spin back to thermal equilibrium. There are multiple possible phonon-involved mechanisms through which this magnetic relaxation can take place. At high temperatures, the relaxation proceeds through the Orbach mechanism via a series of sequential absorptions and emissions of high-energy resonant optical phonons.\cite{C7SC02832F, doi:10.1021/jacs.1c05068} This mechanism shows a characteristic exponential temperature dependence. At low temperatures however, due to the low population of high-energy phonons the magnetic relaxation is instead induced by Raman processes involving low-energy phonons.\cite{doi:10.1021/jacs.1c05068, doi:10.1021/jacs.2c08876} Understanding the interplay between the spin and phonon degrees of freedom and the phonon-assisted relaxation of the magnetic moments in these systems is crucial for the development of high-performance SMMs and their application technologies.

Recent advances have enhanced the understanding of phonon-induced magnetic relaxation in SMMs through first-principles simulations of open quantum systems.\cite{lunghi2017role, doi:10.1021/acs.jpclett.7b00479, 10.1063/5.0017118, doi:10.1021/jacs.1c01410, doi:10.1126/sciadv.abn7880, D4CC03768E, doi:10.1021/jacs.2c08876, doi:10.1021/jacs.1c05068} Most of these studies utilize multireference electronic structure methods such as the complete active space self-consistent field (CASSCF)\cite{roos1980complete, siegbahn1981complete, siegbahn1980comparison} to capture the strongly correlated d- or f-element energy landscapes driving SMM magnetic behavior. However, CASSCF neglects electron correlation outside the active space, which can significantly affect predictions, both quantitatively and qualitatively.

Some popular post-CASSCF methods are N-electron valence perturbation theory to the second-order (NEVPT2)\cite{10.1063/1.1521434, 10.1063/1.1361246, ANGELI2001297}, complete active space second-order perturbation theory (CASPT2)\cite{doi:10.1021/j100377a012, ROOS1996257} and multiconfiguration pair-density functional theory (MC-PDFT).\cite{doi:10.1021/ct500483t, D2SC01022D, annurev:/content/journals/10.1146/annurev-physchem-090419-043839, doi:10.1021/acs.accounts.6b00471} 
The latter offers CASPT2-level accuracy at a significantly reduced computational cost.

Ungur and Chibotaru\cite{https://doi.org/10.1002/chem.201605102} demonstrated that using CASPT2 improves the theoretical prediction of crystal field splitting in lanthanide complexes compared to CASSCF. For an Er-complex they found that CASPT2 corrects the CASSCF-computed crystal field spectrum and magnetic properties, aligning computed values more closely with experimental measurements. Neese and coworkers have reported the crucial role of NEVPT2 in refining the CASSCF-computed SMM properties such as spin-orbit splitting, magnetic anisotropy and spin-Hamiltonian parameters in several transition metal-\cite{doi:10.1021/acs.inorgchem.2c04050, doi:10.1039/D2CP02975H,doi:10.1021/acs.inorgchem.5b01706, Atanasov_2018, atanasov2012modern, singh2018challenges} and lanthanide-based\cite{doi:10.1021/acs.inorgchem.6b00244,jung2017ab} SMMs. The impact of going beyond a CASSCF treatment on spin-phonon relaxation dynamics in SMMs, however, remains underexplored. By altering the energy separation of Kramers doublets and influencing spin-phonon coupling strength, a post-CASSCF method could introduce new relaxation channels or modify relaxation timescales. This study is particularly urgent in the face of common deviations up to one order of magnitude between experiments and simulations. 

\begin{figure}[h!]
\centerline{\includegraphics[width=1.1\textwidth,,height=0.37\textwidth]{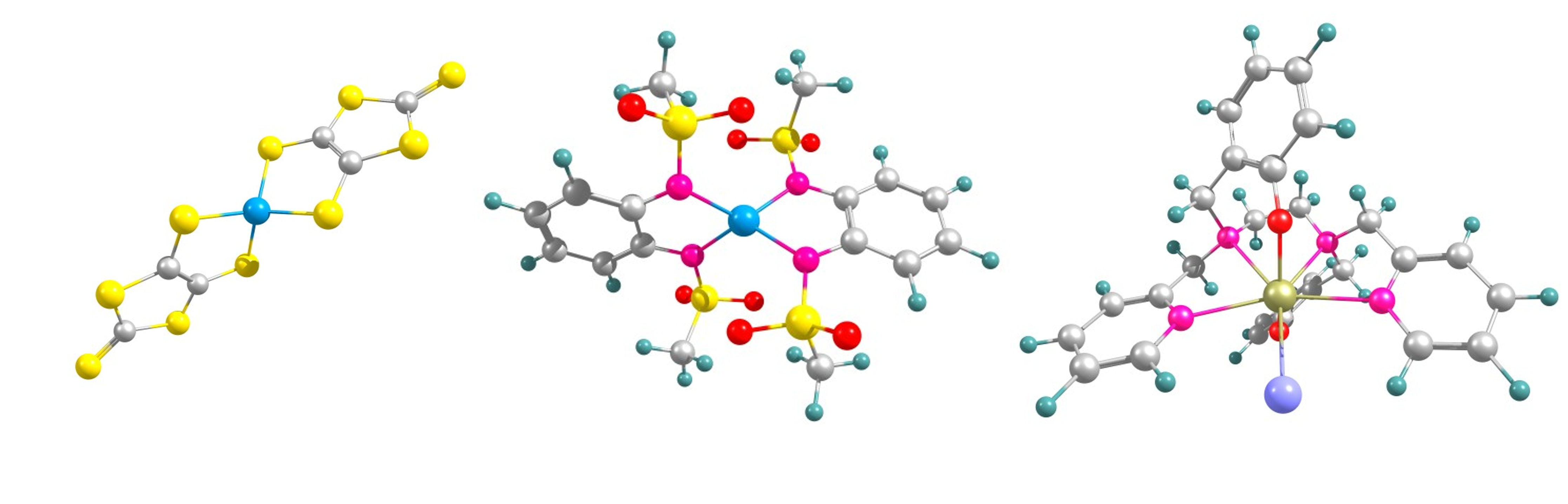}}
\centering
\caption{Structure of the SMMs studied in this work. 
Left: [Co(C\textsubscript{3}S\textsubscript{5})\textsubscript{2}]\textsuperscript{2-} (\textbf{1}) Middle: [CoL\textsubscript{2}]\textsuperscript{2\textsuperscript{-}} (\textbf{2}) Right: [Dy(bbpen)Cl] (\textbf{3}). \textbf{1} and \textbf{2} are shown omitting the counterions, \textbf{3} is neutral. Color codes for atoms:
Co in blue, Dy in golden, S in yellow, C in gray, N in pink, O in red, Cl in light purple, H in light cyan.} 
\label{fig: Molecules Studied}
\end{figure}

This work marks the first systematic application of CASPT2 and MC-PDFT as electronic structure methods for computing spin relaxation in SMMs. The goal is to provide an accuracy superior to CASSCF, but at a similar cost, if MC-PDFT were the method of choice. As case studies, we explore spin-phonon relaxation dynamics in two mononuclear cobalt(II)-based SMMs, namely [Co(C\textsubscript{3}S\textsubscript{5})\textsubscript{2}](Ph\textsubscript{4}P)\textsubscript{2}\cite{doi:10.1021/ic501906z} (\textbf{1}), and [CoL\textsubscript{2}][(HNEt\textsubscript{3})\textsubscript{2}]\cite{rechkemmer2016four} where H\textsubscript{2}L = 1,2-bis (methanesulfonamido)benzene (\textbf{2}), as well as a Dy(III)-based SMM namely [Dy(bbpen)Cl]\cite{doi:10.1021/jacs.6b02638} where H\textsubscript{2}bbpen = N,N'-(bis(2-hydroxybenzyl)
-N,N'-bis (2-methylpyridyl)
ethylenediamine) (\textbf{3}). The structures of \textbf{1}-\textbf{3} (without the counterions) are reported in Fig. \ref{fig: Molecules Studied}. All three SMMs exhibit long spin relaxation times and have been extensively studied before, representing the ideal testbed for determining the importance of electronic correlation for spin-phonon relaxation predictions.

\graphicspath{{Figures/}}

\section{Theory}

\subsection{Electronic Structure Calculations}

The SMMs chosen for this study are multireference in nature. Therefore we employ the state-average complete active space self consistent field method (which we will refer to as CASSCF), to capture the electron correlation inside  the active space (AS). In the following we will refer to it as static correlation. The CASSCF wave function is constructed as a linear combination of all possible configurations within the active space
\begin{equation}
\ket{\Psi_{CASSCF}}=\sum_{n_1n_2...n_L}C_{n_1n_2...n_L}\ket{22...n_1n_2...n_L00} \:,
\end{equation}
where the ket vector represents a specific electronic configuration with "2" being the doubly occupied core orbitals, $n_i$ being the occupation number of the $i^{th}$ active orbital, and, "0" being the unoccupied virtual orbitals. $C_{n_1n_2...n_L}$ are the coefficients for each configuration. The CASSCF energy is expressed as
\begin{equation}
E_{CASSCF}=\sum_{pq}h_{pq}D_{pq}+\sum_{pqrs}g_{pqrs}d_{pqrs}+V_{nn} \:,
\end{equation}
where $p, q, r, s$ are general spatial molecular orbital indices. $h_{pq}$ and $g_{pqrs}$ are the one- and two- electron integrals, $D_{pq}$ and $d_{pqrs}$ are the one- and two-body reduced density matrices respectively and $V_{nn}$ is the sum of the nucleus–nucleus repulsions.

Starting from the reference CASSCF wave function we perform CASPT2 and MC-PDFT calculations. CASPT2 provides a second order perturbation correction to the CASSCF energy.\cite{doi:https://doi.org/10.1002/9780470141526.ch5}
The description of the CASPT2 method can be found in literature.\cite{doi:10.1021/j100377a012, 10.1063/1.462209, https://doi.org/10.1002/qua.23052, 10.1063/1.5097644, BATTAGLIA2023135}

In the MC-PDFT method\cite{doi:10.1021/ct500483t, D2SC01022D, annurev:/content/journals/10.1146/annurev-physchem-090419-043839, doi:10.1021/acs.accounts.6b00471} the classical energy is obtained from the reference CASSCF wave function and then an on-top pair density functional is used to compute the non-classical exchange-correlation energy. The total MC-PDFT energy is expressed as:
\begin{equation}
E_{MC-PDFT}=\sum_{pq}h_{pq}D_{pq}+\sum_{pq>rs}g_{pqrs}D_{pq}D_{rs}+V_{nn}+E_{ot}[\rho,\Pi] \:,
\end{equation}
where the first two terms correspond to the classical energy, and $E_{ot}$ is a functional of the density ($\rho$) and the on-top pair-density ($\Pi$). Different functional forms can be chosen for $E_{ot}$ and the computed energies are dependent on the functional forms. 
The most widely used on-top functional is translated PBE (tPBE)\cite{doi:10.1021/ct500483t, PhysRevLett.77.3865}, with densities and density gradients obtained using the PBE functional form. A hybrid functional called tPBE0 mixes $25\%$ of local exchange with Hartree-Fock exchange.
The accuracy of tPBE and tPBE0 has been shown to be similar to CASPT2  for bond energies\cite{C5SC03321G, doi:10.1021/acs.jctc.6b01102}, spin splitting\cite{10.1063/1.5017132, C6SC05036K, doi:10.1021/acs.jpclett.7b00570},  and excitation energies\cite{doi:10.1021/acs.jctc.5b00456, doi:10.1021/acs.jpclett.5b02773}.  Yet the cost of running an MC-PDFT calculation is comparable to the cost of CASSCF.

While CASPT2 computes a second order correction to the CASSCF energy, usually referred to as dynamic correlation, MC-PDFT uses the CASSCF wave function to compute the total energy with a functional expression. It is thus not formally correct to say that MC-PDFT recovers dynamic correlation. In the following we will thus simply discuss going beyond the CASSCF approximation without distinguishing between static and dynamic correlation, because such a distinction makes sense in the CASPT2 case, but not in the MC-PDFT case.

\subsection{The Model Spin Hamiltonian}
In the field of molecular magnetism, it is common practice to describe the magnetic properties of a system using an effective spin Hamiltonian $\hat{H}_s$ tailored to represent the ground-state magnetic multiplet\cite{10.1063/1.4739763, doi:10.1021/ct900326e}. This subspace of the Hilbert space contains the lowest $2J+1$ states of the system, and its description through an effective spin Hamiltonian offers the primary advantage of simplifying the interpretation of experimental measurements. Furthermore, when the spin Hamiltonian is derived from \textit{ab initio} electronic structure calculations, it becomes possible to exploit this simplified form of the total electronic Hamiltonian $\hat{H}_{el}$ to efficiently compute couplings between the lattice and the spin degrees of freedom\cite{doi:10.1126/sciadv.abn7880}.\\
In the absence of an external magnetic field, the specific form of a generalized model spin Hamiltonian for a single spin system is expressed as\cite{10.1063/1.4739763}
\begin{equation}
    \hat{H_s} = \sum_{l=2(even)}^{2J}\sum_{m=-l}^{l} B_m^l\hat{O}_m^l \:,
    \label{Eq. SH}
\end{equation}
where the operators $\hat{O}_m^l$ are tesseral functions of the total angular momentum operators $\hat{J}$ of rank $l$ and order $m$. $J$ is the total angular momentum quantum number, and $B_m^l$ are the spin Hamiltonian parameters that capture the dependence of the magnetic properties on the electronic structure. We extract these spin Hamiltonian parameters at the equilibrium geometry via a mapping between the matrix elements of $\hat{H}_{el}$ and $\hat{H}_s$, where $\hat{H}_{el}$ is inclusive of spin-orbit coupling (SOC)
%
%
\begin{equation}
    \bra{\hat{J}, m_j}\hat{H}_s\ket{\hat{J}, m_j}=\bra{\hat{J}, m_j}\hat{H}_{el}\ket{\hat{J}, m_j} \:.
    \label{Eq. mapping}
\end{equation}

It is to be noted that for this mapping in Eq. \ref{Eq. mapping}, the spin Hamiltonian and the electronic Hamiltonian must be expressed in a common basis - the spin eigenstates basis $\ket{\hat{J}, m_j}$. Since the spin Hamiltonian is only defined for the lowest $2J+1$ states of the full Hilbert space, the spin basis is obtained by diagonalizing the $(2J+1)\times(2J+1)$ block of $\hat{J}_z$ expressed in \textit{ab initio} basis and opportunely rotated along the easy axes of magnetization of the system.\cite{doi:10.1021/acs.jctc.3c01130} The \textit{ab initio} basis is obtained by diagonalizing the electronic Hamiltonian $\hat{H}_{el}$ in the spin-free basis. 
The \textit{ab initio} basis constructed from these methods are thus different, by virtue of which the spin eigenstates of $\hat{J}_z$ are also different for different methods. As a result, the new spin eigenstates basis sets are different for CASSCF, CASPT2 and MC-PDFT.
The resulting spin-Hamiltonian parameters extracted from Eq. \ref{Eq. mapping} are therefore also different for different methods which in turn should be reflected in the computed relaxation time. In this and the following sections, we use $J$ to denote the total angular momentum of the system. However, it is important to note that for Co(II)-based compounds, the orbital angular momentum $L$ is quenched, resulting in $J=S$. In this case, the Hamiltonian in Eq. \ref{Eq. SH} only contains the term with $l=2$, which is equivalent to the standard zero-field splitting Hamiltonian
\begin{equation}
    \hat{H}_s = \vec{\mathbf{S}} \cdot \mathbf{D} \cdot \vec{\mathbf{S}} \:.
\end{equation}

\subsection{Relaxation Dynamics}

In order to simulate the phonon-induced spin-relaxation dynamics the total Hamiltonian of the entire system is constructed by adding the three contributing terms coming from the spin subsystem, the phonon system, and the contribution due to the interaction between these subsystems
\begin{equation}
    \hat{H}=\hat{H}_s+\hat{H}_{ph}+\hat{H}_{s-ph}\:,
    \label{Eq: Total H}
\end{equation}
where $\hat{H_s}$ is the model spin Hamiltonian as introduced in Eq. \ref{Eq. SH}. $\hat{H}_{ph}$ is the component that describes the phonon modes of the system, approximated as a sum of harmonic oscillators
\begin{equation}
    \hat{H}_{ph}=\sum_{\alpha q} \hbar\omega_{\alpha q}(\hat{n}_{\alpha q}+\frac{1}{2})\:,
    \label{Eq. harmonic}
\end{equation}
where $\hat{n}_{\alpha q}$ is the phonon number operator for a particular phonon mode with frequency $\omega_{\alpha q}$. The summations run over the phonon mode indices $\alpha$ as well as the reciprocal lattice vectors $q$. Since we consider phonons only at the Gamma point i.e. the center of the Brillouin zone ($q=(0,0,0)$), we drop the index $q$ for the rest of the discussion.

The last component $\hat{H}_{s-ph}$ is the spin-phonon coupling term, which captures the intensity or strength of the interaction of the electronic subsystem with the weakly coupled phonon bath. During the crystal vibrations, the effect of the slight changes in nuclear coordinates on the magnetic properties is quite small and can be modeled as perturbations. Therefore, to obtain the coupling term under the weak coupling approximation the spin Hamiltonian can be expressed as a Taylor expansion around the equilibrium geometry with respect to the nuclear displacements.\cite{Lunghi2023} 
\begin{equation}
    \hat{H}_s(t)=(\hat{H}_s)_0+\sum_{\alpha}\left(\frac{\partial\hat{H}_s}{\partial Q_{\alpha}}\right)_0\hat{Q}_{\alpha}(t)+\frac{1}{2} \sum_{\alpha}\sum_{\beta}\left(\frac{\partial^2\hat{H}_s}{\partial Q_{\alpha}\partial Q_{\beta}}\right)_0 \hat{Q}_{\alpha}(t)\hat{Q}_{\beta}(t)+... \:,
    \label{Eq. Taylor series}
\end{equation}
where the zeroth order term $(\hat{H}_s)_0$ corresponds to the spin Hamiltonian at equilibrium geometry which appears as the first term in Eq. \ref{Eq: Total H}, and the higher-order terms that explicitly depend on time describe the coupling Hamiltonian $\hat{H}_{s-ph}$. As discussed in the previous section, the spin Hamiltonian parameters are highly sensitive to even slight nuclear distortions. In fact, it is at the heart of spin-phonon coupling simulation - the coupling coefficients in Eq. \ref{Eq. Taylor series} are obtained as first-order derivatives of the crystal field parameters $B_m^l$ with respect to the molecular Cartesian coordinates. They are then transformed into the crystal coordinate basis according to the following relation\cite{Lunghi2023, doi:10.1021/jacs.2c08876}
\begin{equation}
    \frac{\partial B_m^l}{\partial Q_\alpha}=\sum_a^{3N}\sqrt{\frac{\hbar}{2\omega_\alpha m_a}}L_{\alpha a}\biggl(\frac{\partial B_m^l}{\partial R_a}\biggr) \:,
     \label{Eq: nomalcoord}
\end{equation}
where $Q_{\alpha}$ is the displacement vector corresponding to the normal mode $\alpha$ with angular frequency $\omega_\alpha$, $N$ is the number of atoms in the unit cell, $L_{\alpha a}$ is the Hessian matrix eigenvector, and $R_a$ is the Cartesian degree of freedom. With the knowledge of the coupling coefficients, the relaxation rate for the transition between two spin eigenstates $\ket{a}$ and $\ket{b}$ is determined under the Born-Markov approximation using the time evolution of the density matrix of the open quantum system. For the relaxation involving a single resonant phonon the rate is\cite{doi:10.1021/jacs.2c08876, doi:10.1126/sciadv.abn7880}
\begin{equation}
    W_{ba}^{1-ph}=\frac{2\pi}{\hbar^2}\sum_\alpha|\bra{b}\biggl(\frac{\partial \hat{H}}{\partial Q_\alpha}\biggr)\ket{a}|^2G^{1-ph}(\omega_{ba},\omega_{\alpha}) \:,
    \label{Eq: transition_rate Orbach}
\end{equation}
where $\hbar\omega_{ab} = E_a - E_b$ and the function $G^{1-ph}$ is expressed as
\begin{equation}
    G^{1-ph}(\omega,\omega_{\alpha}) = \delta(\omega-\omega_{\alpha})
    \Bar{n}_{\alpha}+\delta(\omega+\omega_{\alpha})(\Bar{n}_{\alpha}+1) \:,
    \label{G1ph}
\end{equation}
where $\Bar{n}_{\alpha}=\frac{1}{exp(\hbar\omega_{\alpha}/k_BT)-1}$ is the Bose-Einstein distribution for the thermal population of phonons, $k_B$ is the Boltzmann constant and the first and second Dirac $\delta$ functions enforce the energy conservation during the absorption and emission of phonons, respectively. The temperature dependence of spin-phonon relaxation rate arises through this thermal population of phonons and their energy distribution.

For the two-phonon Raman relaxation the transition rate is\cite{doi:10.1021/jacs.2c08876} 
\begin{equation}
    W_{ba}^{2-ph}=\frac{2\pi}{\hbar^2}\sum_{\alpha\beta}|T_{ab}^{\alpha\beta,+}+T_{ab}^{\beta\alpha,-}|^2G^{2-ph}(\omega_{ba},\omega_{\alpha},\omega_{\beta}) \:.
    \label{Eq: transition_rate Raman}
\end{equation}
where the function $G^{2-ph}$ accounts for the process involving simultaneous absorption and emission of two phonons and is expressed as
\begin{equation}
    G^{2-ph}(\omega,\omega_{\alpha}, \omega_{\beta}) = \delta(\omega-\omega_{\alpha}+\omega_{\beta})
    \Bar{n}_{\alpha}(\Bar{n}_{\beta}+1) \:,
    \label{G2ph}
\end{equation}
where
\begin{equation}
    T_{ab}^{\alpha\beta,\pm}=\sum_k\frac{\bra{a}(\frac{\partial \hat{H}}{\partial Q_\alpha})\ket{k}\bra{k}(\frac{\partial \hat{H}}{\partial Q_\alpha})\ket{b}}{E_k-E_b\pm\hbar\omega_{\beta}} \:.
    \label{Eq. transition_probability Raman}
\end{equation}
The envelope of spin excited states, $\ket{k}$, that appears in Eq. \ref{Eq. transition_probability Raman} is commonly referred to as a virtual state to highlight the fact that such states do not get populated during Raman relaxation.\cite{doi:10.1021/jacs.2c08876} Finally, the relaxation time ($\tau$) is obtained by diagonalizing the matrix $W_{ba}^{n-ph}$ and taking the inverse of the smallest non-zero eigenvalue. $\tau$ obtained from $W_{ba}^{1-ph}$ provides the Orbach contribution to the total relaxation time ($\tau_{Orbach}$), whereas $\tau$ obtained from $W_{ba}^{2-ph}$ provides the Raman contribution to the total relaxation time ($\tau_{Raman}$). Finally, the total spin relaxation time $\tau$ due to the combined effect of both Orbach and Raman relaxation mechanisms can be obtained as $\bigg(\frac{1}{\tau}\bigg)^{-1}=\bigg(\frac{1}{\tau_{Orbach}}+\frac{1}{\tau_{Raman}}\bigg)^{-1}$.

\section{Computational Methods}

For the three compounds investigated in this study, periodic density functional theory (pDFT) cell optimizations and $\Gamma$-point phonon calculations were previously performed by Mondal et al.\cite{doi:10.1021/jacs.2c08876} using the CP2K software\cite{10.1063/5.0007045}. For these calculations, the experimental X-ray structures were used as starting points for geometry optimization and the PBE functional\cite{PhysRevLett.77.3865} with inclusion of the D3 dispersion correction\cite{10.1063/1.3382344} was employed. We used the same optimized molecular geometries and phonons modes for this work. The electronic structure calculations were performed using OpenMolcas version 24.02\cite{https://doi.org/10.1002/jcc.24221}. 
The second-order Douglas–Kroll–Hess (DKH) Hamiltonian was used to account for the scalar relativistic effects, along with all-electron ANO-RCC basis sets contracted to polarized triple-$\zeta$ quality (ANO-RCC-VTZP) for Co and Dy, and double-$\zeta$ quality (ANO-RCC-VDZ) for rest of the atoms. An active space (AS) of seven electrons in the five 3d orbitals (7e, 5o) is used for the Co(II)-compounds. For the Dy(III)-compound two different active spaces were considered. The minimal AS consists of nine electrons in the seven 4f orbitals (9e, 7o). The second active space includes also a second shell of f orbitals (9e, 14o). For compounds \textbf{1} and \textbf{2} a state-average CASSCF calculation was performed over all possible (10) quartet states. For compound \textbf{3}, a state average over all possible (21) sextet states was performed with the (9e, 7o) AS. CASPT2 and MC-PDFT calculations were carried out using these reference CASSCF wave functions. For \textbf{1} and \textbf{2} SOC among all the spin-free quartet states (among all the spin-free sextet states in case of \textbf{3}) were incorporated through the atomic mean-field integral (AMFI) approximation implemented in the restricted active space state interaction (RASSI) module of OpenMolcas\cite{MALMQVIST2002230}. Doublets for the Co(II) systems (doublets and quartets for the Dy(III) system) were excluded in the SOC calculations due to significant energy separation from the low-lying quartets (sextets for the Dy(III) system). The translated “on-top” PBE functional (“tPBE”)\cite{doi:10.1021/ct500483t} was used to compute the MC-PDFT energies.\\
From each \textit{ab initio} calculation, the crystal field parameters $B_m^l$, which fully define the spin Hamiltonian of the the system  (Eq. \ref{Eq. SH}) can be computed. These parameters are derived from the electronic Hamiltonian, $\hat{H}_{el}$, expressed in the spin-free basis, together with the spin and orbital angular momentum matrices $\hat{S}_i$ and $\hat{L}_i$, with $i=x,y,z$. The mapping shown in Eq. \ref{Eq. mapping} is performed with the $get\_CF.x$ routine distributed with the MolForge software\cite{doi:10.1126/sciadv.abn7880} (freely available at github.com/LunghiGroup/MolForge). The MolForge software also allows the calculation of the Orbach and Raman relaxation times by implementing Eqs. \ref{Eq: transition_rate Orbach} and \ref{Eq: transition_rate Raman}. To set up the calculation, firstly the three terms of the Hamiltonian in Eq. \ref{Eq: Total H} must be provided. The ground-state electronic part of the Hamiltonian, $\hat{H}_{s}$, is expressed as a list of crystal field parameters $B_m^l$, generated by the routine $get\_CF.x$.
The phonon bath, $\hat{H}_{ph}$, is constructed from the Hessian matrix obtained with pDFT. The code internally extracts the Hessian matrix eigenvectors $L_{\alpha a}$ and their corresponding frequencies $\omega_{\alpha}$, which are used in Eq. \ref{Eq: nomalcoord}.\\
The spin-phonon coupling matrix elements in Cartesian coordinates are obtained through numerical differentiation of the crystal field parameters with a step of $\pm 0.01$ \r{A}. These values are provided in input to MolForge which internally transforms them in the basis of the crystal coordinates by using Eq. \ref{Eq: nomalcoord}.\\
In addition to the total Hamiltonian, the user must specify the Euler angles linking the molecular reference framework (defined by the raw input atomic positions) to the framework where the $z$-axis aligns with the magnetic easy axis of the system. This axis is defined as the principal eigenvector of the matrix $gg^T$, where $g$ is the $3$x$3$ $g$-tensor of the first Kramers doublet and $g^T$ its transpose. Defining the Euler angles is critical for applying a small magnetic field of $\sim 0.01$ Tesla along the easy axis of magnetization when computing Raman relaxation times, as this decouples the coherence and population density matrix elements\cite{doi:10.1126/sciadv.abn7880}.\\
Finally, for a user-defined set of temperatures the programs outputs the values of $\tau_{Orbach}$ and $\tau_{Raman}$. The role of temperature in these simulations is to modify the Bose-Einstein population term $\Bar{n}_{\alpha}$, which enters the expressions of the transition rates through $G^{1-ph}$ and $G^{2-ph}$ (Eq. \ref{G1ph} and Eq. \ref{G2ph}, respectively).


\section{Results and Discussion}

\subsection{\textbf{A.} [Co(C\textsubscript{3}S\textsubscript{5})\textsubscript{2}](Ph\textsubscript{4}P)\textsubscript{2} (\textbf{1})}

The ground state of the complex \textbf{1} is a $S=3/2$ spin-state, giving rise to 2 sets of Kramers doublets separated by 282 cm$^{-1}$
at the CASSCF level. The CASPT2 and MC-PDFT methods slightly open up the gap to 303 cm$^{-1}$ and 309 cm$^{-1}$, respectively. (See Table \ref{tab: En KD mol_1})

\begin{table}[htbp]
  \begin{center}
  \caption{\textbf{Energies of the lowest Kramers doublets (in cm$^{-1}$) for complex \textbf{1}}}
  \label{tab: En KD mol_1}
  \begin{tabular}{c c c c}
\hline
  \textbf{States} & CASSCF & CASPT2 & MC-PDFT \\
  \hline
    KD$_0$ & 0 &
    0 & 0
     \\ 
    KD$_1$ & 282 &
    303 & 309 \\ 
    \hline
    \end{tabular}
    \end{center}
 \end{table}

Importantly, we observed that the numerical derivatives of the crystal field parameters i.e. the spin-phonon coupling coefficients ($\frac{\partial B_m^l}{\partial Q_\alpha}$) are quite different between CASSCF and CASPT2 methods, whereas they are relatively consistent between the CASSCF and MC-PDFT methods as can be seen from the parity plots in Fig. S1 (a) and (b) respectively. On the other hand, the coupling coefficients differ significantly between the CASPT2 and MC-PDFT methods, although they provide very similar energy gaps between the KDs (Fig. S2). This suggests that the coupling strengths do not have a one to one correspondence with the excitation energies from the ground to excited KDs.

\begin{figure}[h!]
\centerline{\includegraphics[width=0.8\textwidth,,height=0.55\textwidth]{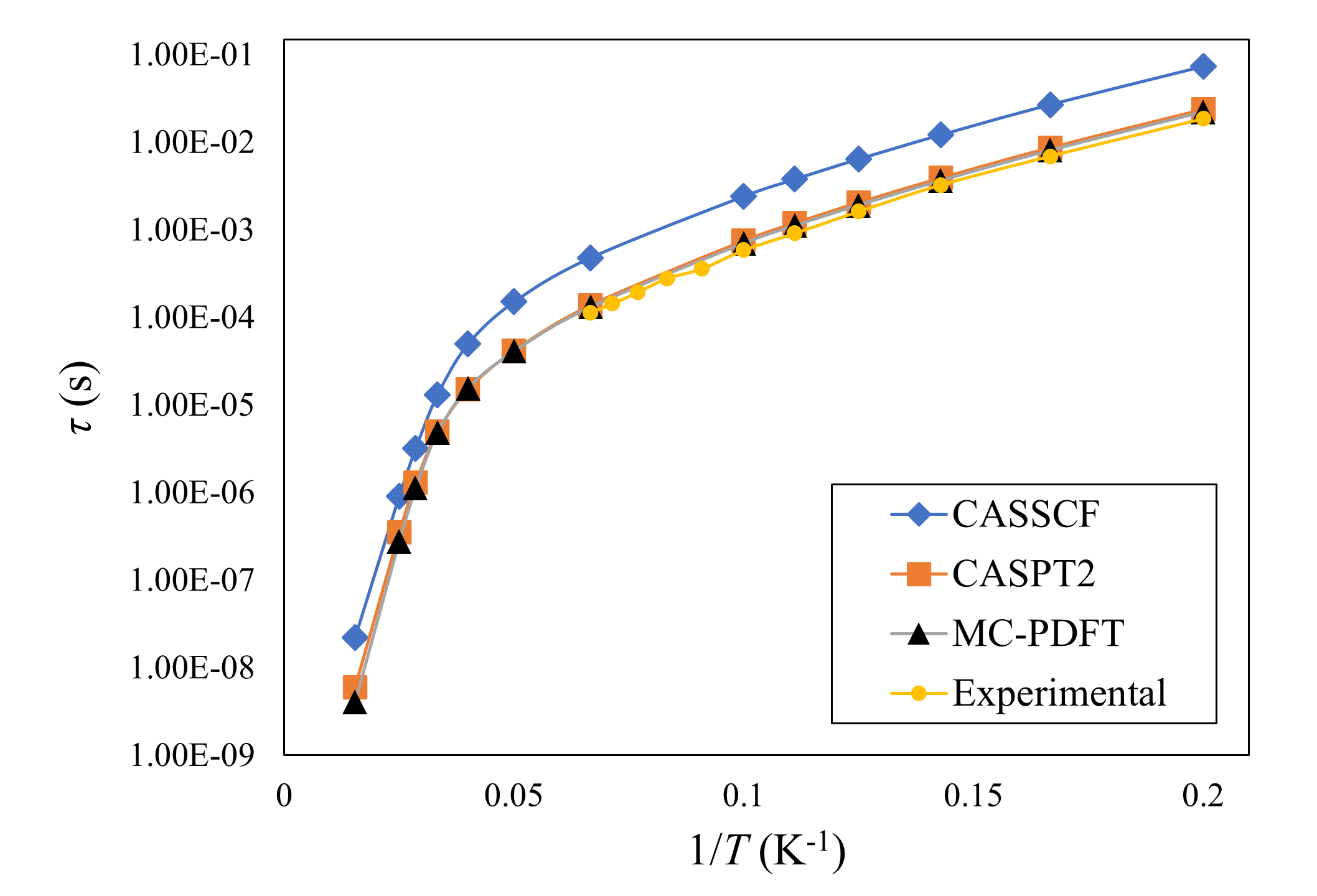}}
\centering
\caption{Total spin relaxation time ($\tau$) as a function of 1/\textit{T} for complex \textbf{1} obtained from different methods. Experimental data is taken from Ref. \cite{doi:10.1021/ic501906z}}
\label{fig: spin-relaxation time mol_1}
\end{figure}

We then computed the Orbach and Raman relaxation rates using Eq.\ref{Eq: transition_rate Orbach} and Eq.\ref{Eq: transition_rate Raman} respectively, at different temperatures. The corresponding relaxation times ($\tau_{Orbach}$ and $\tau_{Raman}$) are obtained by taking the inverse of the rates. The total relaxation times $\tau$ at different temperatures computed with CASSCF, CASPT2, and MC-PDFT are reported in Table S1. The temperature dependence of $\tau$ obtained from these three methods along with the available experimental data for compound \textbf{1} is shown in Fig. \ref{fig: spin-relaxation time mol_1}. 

\graphicspath{{Figures/}}

Although the overall trend in $\tau$ follows the experimental trend for all three methods, $\tau$ is overestimated by an order of magnitude at the CASSCF level compared to the experimental data, while CASPT2 and MC-PDFT provide very similar spin-relaxation time, and they both agree well with the experimental data. We could compare the relaxation time with the experimental data only in the low temperature region due to the unavailability of experimental data at high temperatures between 20 K to 65 K ($1/\textit{T}$ between $\sim$ 0.07 K\textsuperscript{-1} to $\sim$ 0.02 K\textsuperscript{-1}). An enlarged view of Fig. \ref{fig: spin-relaxation time mol_1} is provided in Fig. S3, offering a clearer illustration of the agreement between CASPT2 and MC-PDFT predictions with the experimental results.

Despite differences in spin-phonon coupling coefficients between the two post-CASSCF methods, both yield almost identical spin relaxation times for complex \textbf{1}. The reason is that, besides the coupling coefficient, other factors contribute to the rate expressions (Eq. \ref{Eq: transition_rate Orbach} and Eq. \ref{Eq: transition_rate Raman}), namely the spin eigenstates. The shape of time vs 1/\textit{T} curves thus reflects all of them.

\subsection{\textbf{B.} [CoL\textsubscript{2}][(HNEt\textsubscript{3})\textsubscript{2}] (\textbf{2})}

Similar to complex \textbf{1} the ground state of the complex \textbf{2} has spin $S=3/2$. The Kramers doublets are separated by  207 cm$^{-1}$ at CASSCF, 254 cm$^{-1}$ at CASPT2 and 191 cm$^{-1}$ at MC-PDFT levels. (Table \ref{tab: En KD mol_2})
The spin-phonon parameters are reasonably consistent across all three methods - as illustrated in Fig. S4 and S5. Like for complex \textbf{1}, this suggests that the coupling strengths do not have a direct correspondence with the KD-excitation energies. 

\begin{table}[htbp]
  \begin{center}
  \caption{\textbf{Energies of the lowest Kramers doublets (in cm$^{-1}$) for complex \textbf{2}}}
  \label{tab: En KD mol_2}
  \begin{tabular}{c c c c}
\hline
  \textbf{States} & CASSCF & CASPT2 & MC-PDFT \\
  \hline
    KD$_0$ & 0 &
    0 & 0
     \\ 
    KD$_1$ & 207 &
    254 & 191 \\ 
    \hline
    \end{tabular}
    \end{center}
 \end{table}

\graphicspath{{Figures/}}
\begin{figure}[h!]
\centerline{\includegraphics[width=0.8\textwidth,height=0.55\textwidth]{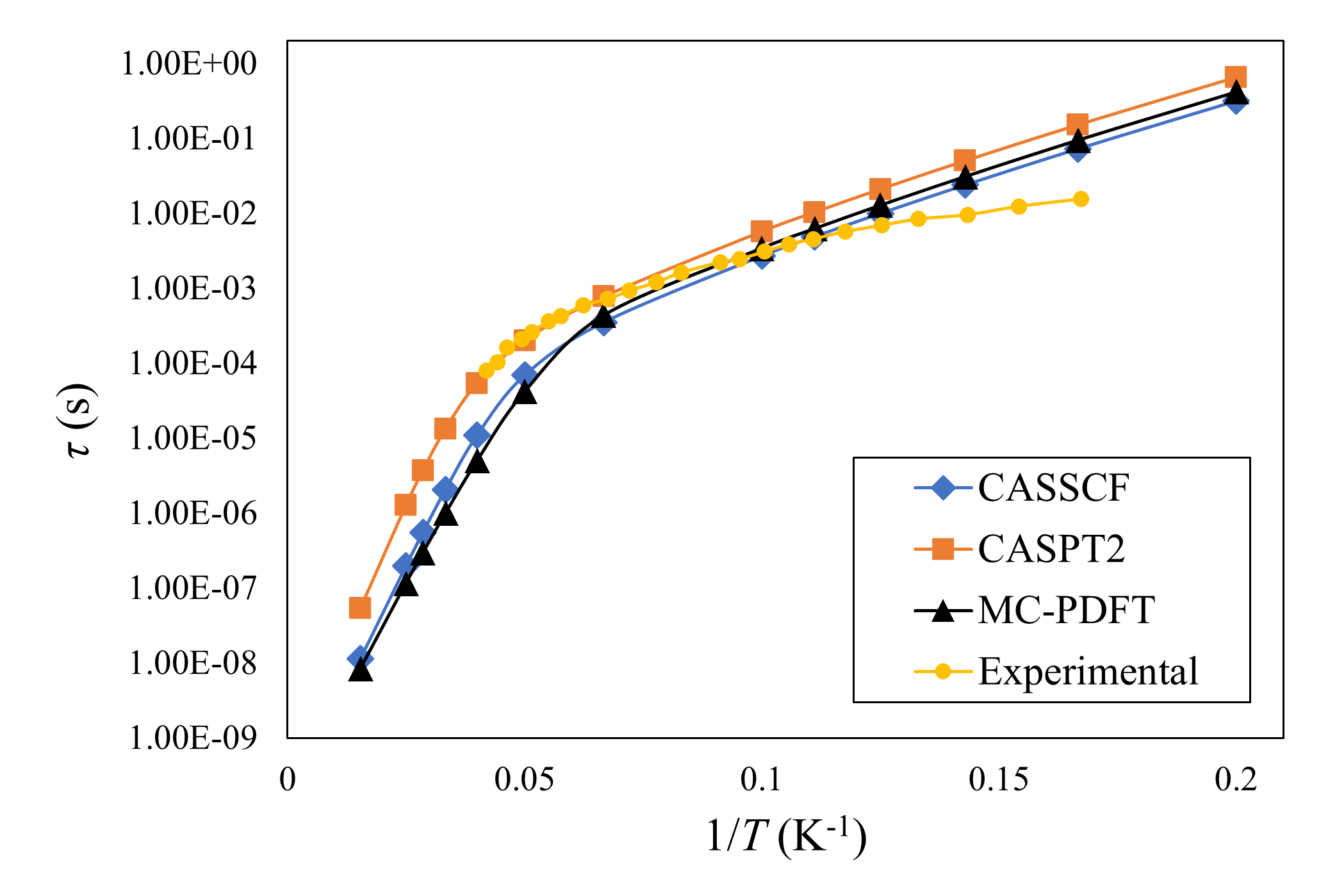}}
\centering
\caption{Total spin relaxation time ($\tau$) as a function of 1/\textit{T} for complex \textbf{2} obtained from different methods. Experimental data is taken from Ref. \cite{rechkemmer2016four}}
\label{fig: spin-relaxation time mol_2}
\end{figure}

Fig. \ref{fig: spin-relaxation time mol_2} shows that $\tau$ obtained from CASPT2 overlaps with the experimental relaxation times in the temperature range between 15 K and 20 K ($1/\textit{T}$ between 0.07 K\textsuperscript{-1} to 0.04 K\textsuperscript{-1}), whereas CASSCF and MC-PDFT values are lower. An enlarged view of Fig. \ref{fig: spin-relaxation time mol_2} is provided in Fig. S6, clearly illustrating the agreement between CASPT2 and experimental results. Between 5K to 10K ($1/\textit{T}$ between 0.2 K\textsuperscript{-1} to 0.1 K\textsuperscript{-1}), the deviation of all the computed $\tau$ values from the experimental data arises due to the lack of consideration of the quantum tunneling of magnetization (QTM) mechanism of spin relaxation in our simulation - which appears to significantly contribute to the experimentally observed total spin relaxation. At very high temperatures between 20 K to 65 K ($1/\textit{T}$ from 0.05 K\textsuperscript{-1} to 0.01 K\textsuperscript{-1}) the relaxation times predicted by CASSCF and MC-PDFT are in close agreement, while CASPT2 estimates slightly longer relaxation times. This trend aligns with expected behavior, although experimental data is unavailable for this temperature range.

\subsection{\textbf{C.} [Dy(bbpen)Cl] (\textbf{3})}

\begin{figure}[h!]
\centerline{\includegraphics[width=0.70\textwidth,height=0.70\textwidth]{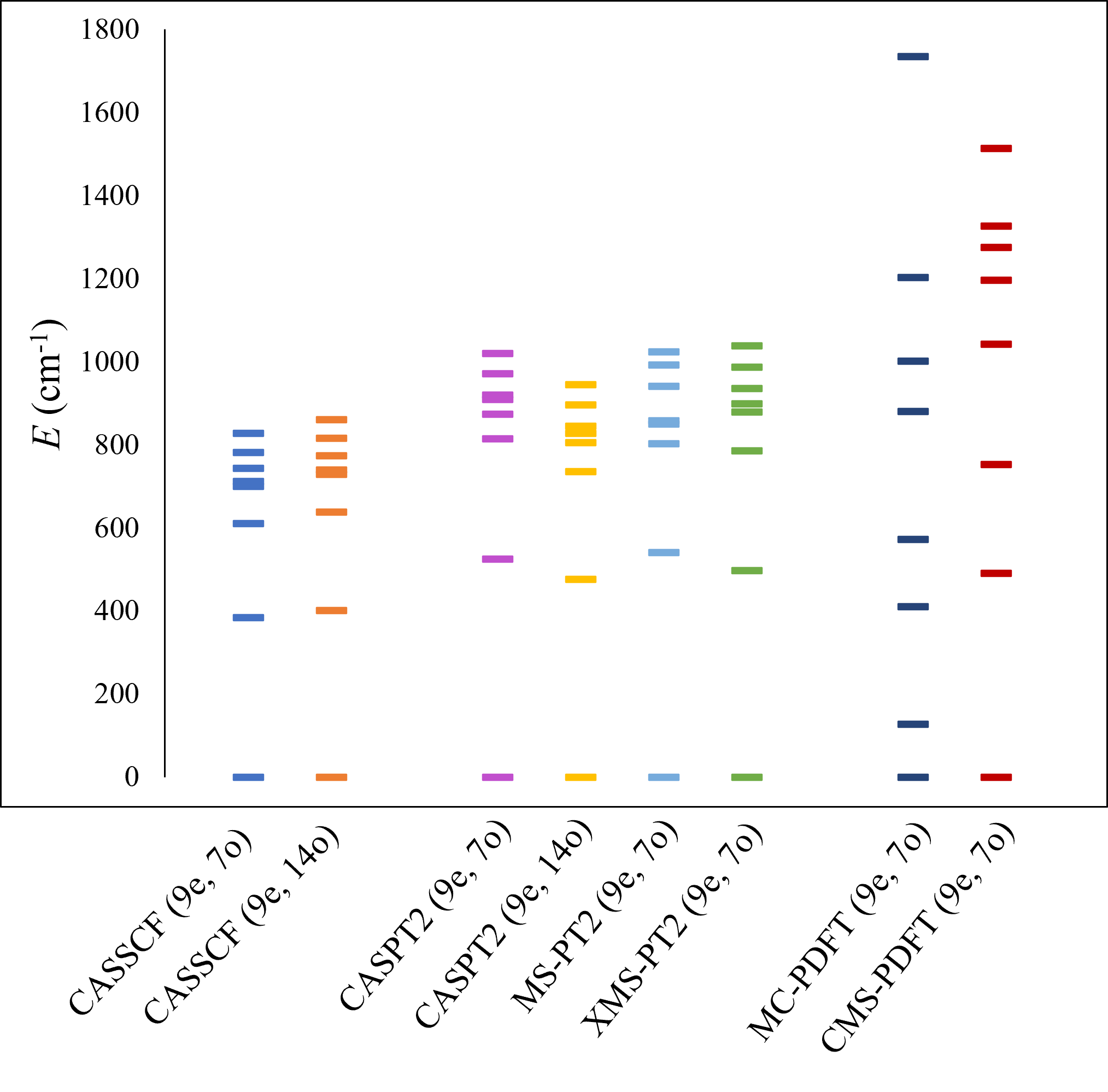}}
\centering
\caption{Energy spacings among the ground and excited Kramers doublets obtained from different methods for compound \textbf{3}. Active spaces employed for the calculations are mentioned in parenthesis.}
\label{fig: En KDs mol_3}
\end{figure}

Complex \textbf{3} exhibits a ground state multiplet $^6H_{15/2}$ with total angular momentum $J=15/2$ giving rise to 8 Kramers doublets. The energy spacing among these Kramers doublets obtained from different methods are shown in Fig. \ref{fig: En KDs mol_3}. The absolute energies of all the KDs computed by these methods are provided in Table S3.

\graphicspath{{Figures/}}

\begin{figure}[h!]
\centerline{\includegraphics[width=0.75\textwidth,height=0.55\textwidth]{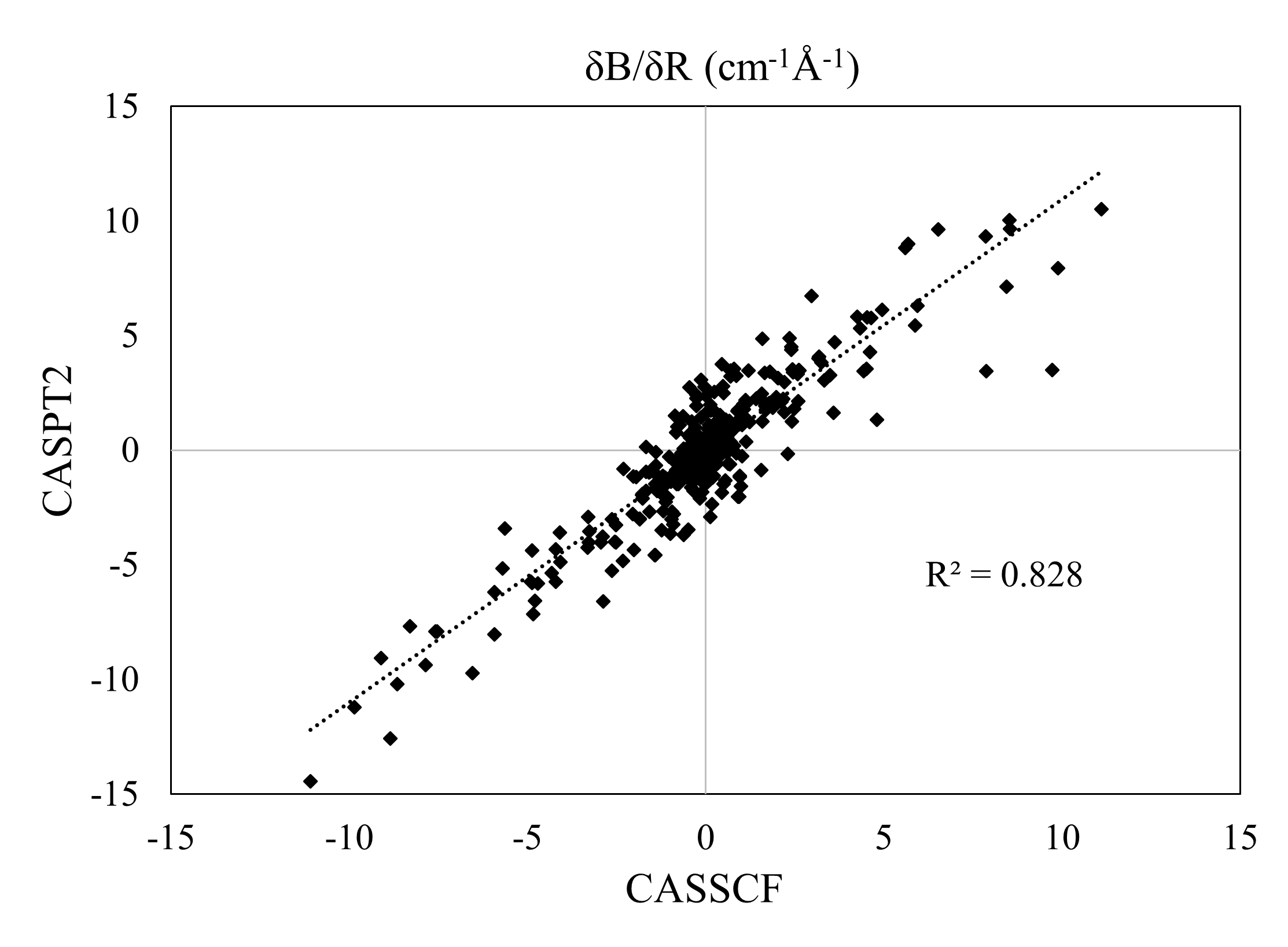}}
\centering
\caption{Parity plots comparing the numerical derivatives of the crystal field parameters computed at CASSCF and CASPT2 levels for compound \textbf{3}.}
\label{fig: Parity plot mol_3}
\end{figure}

We begin by focusing on the results obtained using the smaller active space (9e, 7o). Similar to complex \textbf{1} and \textbf{2}, CASPT2 opens up the gap between the ground and first excited KD. The CASPT2 energy difference ($\Delta$KD$_0^7$) between the ground KD (KD$_0$) and the 7$^{th}$ excited KD (KD$_7$) is 1020 cm$^{-1}$ compared to the CASSCF value of 827 cm$^{-1}$. On the other hand, with MC-PDFT, the 1st excited KD (KD$_1$) is very close to KD$_0$, unlike CASSCF and CASPT2, and $\Delta$KD$_0^7$ is high in energy, 1734 cm$^{-1}$. 
Due to the possible strong interaction among the closely spaced KDs, we also investigated the performance of the multistate versions of these methods - namely, multistate\cite{FINLEY1998299} (MS), extended multistate\cite{10.1063/1.3596699, 10.1063/1.3633329} (XMS) CASPT2, and compressed multistate PDFT (CMS-PDFT)\cite{doi:10.1021/acs.jctc.0c00908}. MS-CASPT2 and XMS-CASPT2 give similar energy levels to CASPT2 (Fig. \ref{fig: En KDs mol_3}). On the other hand, CMS-PDFT provides a much more consistent energy spacing among the lowest KDs as compared to the single-state MC-PDFT - although $\Delta$KD$_0^7$ is still very large (1513 cm$^{-1}$).
This behavior has been previously detected for lanthanide and actinide compounds with dense energy levels\cite{doi:10.1021/acs.jpca.3c05803} and new state interaction formulations of PDFT\cite{doi:10.1021/acs.jctc.3c00207} are under investigation. In the following we will only discuss the CASPT2 results and compare them with previously reported CASSCF results.
\cite{doi:10.1021/jacs.6b02638} The spin-phonon coupling coefficients are rather consistent between these two methods. (Fig. \ref{fig: Parity plot mol_3})

\graphicspath{{Figures/}}

\begin{figure}[h!]
\centerline{\includegraphics[width=0.8\textwidth,,height=0.55\textwidth]{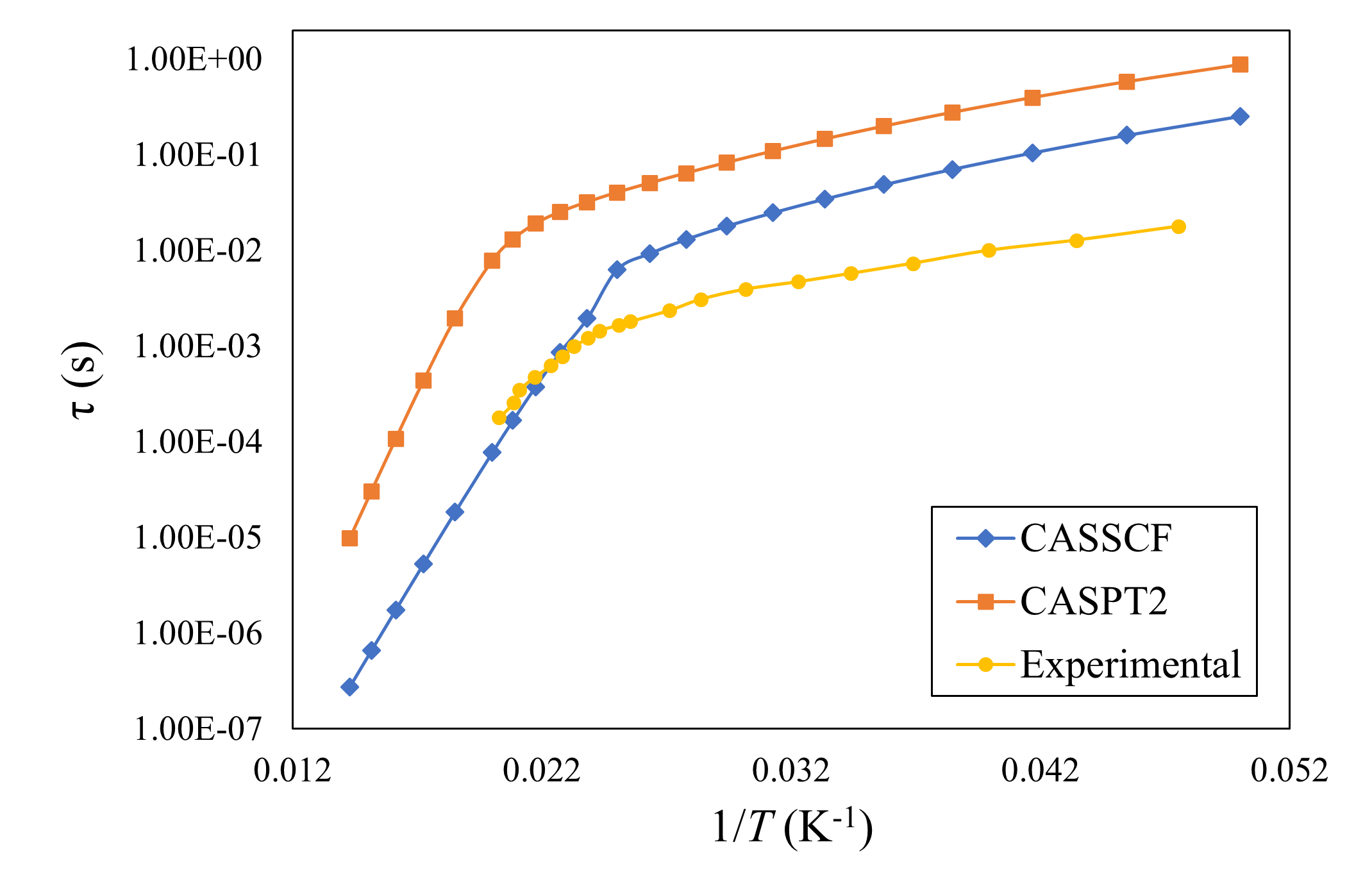}}
\centering
\caption{Total spin relaxation time ($\tau$) as a function of 1/\textit{T} for complex \textbf{3} obtained from different methods. Experimental data is taken from Ref.\cite{doi:10.1021/jacs.6b02638}}
\label{fig: spin-relaxation time mol_3}
\end{figure}

The overall trends of CASSCF and CASPT2 total relaxation time vs. temperature reproduce the experimental trend (Fig. \ref{fig: spin-relaxation time mol_3}). However, CASSCF and CASPT2 overestimate the experimental data by one and two orders of magnitude, respectively. One difference between the two methods is that CASPT2 predicts too high Kramers doublet energies compared to CASSCF. It has been previously reported for lanthanide systems that when a small active space is used CASPT2 may predict too high crystal field splitting. \cite{https://doi.org/10.1002/chem.201605102} Expanding the AS by means of a second shell of f orbitals may counteract this CASPT2 effect  and provide more accurate crystal field splittings. For the equilibrium geometry we tested the CASPT2 performance using the (9e, 14o) AS. We used as a guess the SA-CASSCF(9e, 7o) wave function, and considered an averaging over the same number of states (21). We indeed observed a significant reduction in the crystal field splitting of the ground electronic state (Fig. \ref{fig: En KDs mol_3}). However, the increase in computational cost associated with the larger AS restricts us to use it for the study of spin-relaxation dynamics - especially since the CASSCF and CASPT2 need to be performed for all the distorted structures to compute the numerical differentiation of the crystal field Hamiltonian. 

\graphicspath{{Figures/}}

\newpage
\section{Discussion and Conclusions}

CASSCF methods have been extensively used to simulate spin relaxation time as a function of the temperature of the phonons bath and have been shown to reproduce experimental trends.\cite{doi:10.1126/sciadv.abn7880, doi:10.1021/jacs.2c08876, doi:10.1021/jacs.3c06015} However, the systematic exploration of multiple molecules has shown that quantitative inaccuracies persist.\cite{doi:10.1021/jacs.2c08876} In particular, it seems not currently possible to confidently rank the relaxation rate of different compounds unless they differ by at least one order of magnitude. The present study presents the first systematic exploration of multireference post-CASSCF methods, CASPT2 and MC-PDFT, for the simulations of spin relaxation in mononuclear single-molecule magnets and results show that the dynamical electron correlation beyond the active space is critical in explaining the discrepancies previously observed. 

CASPT2 systematically improves the agreement between experiments and simulations and achieves quantitative accuracy for both the Co compounds studied. Interestingly, this is observed both in cases where CASSCF underestimates or overestimates experimental results by one order of magnitude, suggesting that CASPT2 is able to capture non-trivial correlations between chemical structure, zero-field splitting and spin-phonon coupling. MC-PDFT on the other hand matches the accuracy of CASPT2 for System \textbf{1}, but retains the same level of error of CASSCF for System \textbf{2}. A detailed analysis of all the computed quantities determining spin relaxation, i.e. the Hamiltonian egivenvalues, Hamiltonian eigenstates, and spin-phonon coupling coefficients, highlights the absence of a simple explanation for CASSCF's deficiencies, and that the improvement of the spin relaxation time at the CASPT2 level is a consequence of overall simultaneous better accuracy for all these quantities. Although larger benchmarks will be necessary to prove the generality of this important result, CASPT2 appears to achieve the goal of quantitative and systematic predictions for Co-based mononuclear compounds. 

The situation is drastically different for the Dy compound (System \textbf{3}), where all computational methods deviate significantly from experiments and discrepancies span one to two orders of magnitude. Interestingly, our results suggest that the overall high-level and systematic accuracy observed in the prediction of spin-phonon relaxation times in Dy-compounds with CASSCF might be partially due to a cancellation of errors. Indeed, here we have shown that the inclusion of correlation beyond the active space through CASPT2 drastically decreases the accuracy of predictions. The use of a larger active space seems to alleviate this issue, but full convergence of results with respect to the active space size could not be achieved without incurring extensive computational costs.

The difficulty in converging the size of the active space for the Dy compound mostly comes from the large expense of computing spin-phonon coupling coefficients, which requires at least six times the number of atoms CASPT2 calculations. Numerical approaches to reduce the expense of multi-reference calculations or the number of single-point calculations to compute spin-phonon coupling are an urgent necessity. On the former front, MC-PDFT stands out as a promising route, but our results show that further development is required to consistently achieve CASPT2 levels of accuracy. In terms of lowering the number of calculations to estimate spin-phonon coupling two routes have been recently pursued. The use of analytical gradients has been proposed,\cite{doi:10.1021/acs.jctc.2c00611} but it currently lacks the contribution of spin-orbit coupling derivatives, which has been shown to lead to errors for both Co and Dy compounds,\cite{doi:10.1021/acs.jctc.3c01130} including System (\textbf{3}). To the best of our knowledge, the implementation of the analytical gradients of the spin-orbit coupling operators has never been pursued but represents an interesting avenue of investigation. On the other hand, analytical gradients are not available for all methods and all codes, calling for alternative numerical strategies. In this regard, machine learning methods offer a very promising alternative.\cite{lunghi2022computational} Seminal steps have been pursued in this direction and savings up to 80\% in the computation of phonons and spin-phonon coupling have been demonstrated.\cite{briganti2024machinelearningframeworkacceleratingspinlattice} We anticipate that the combination of these data-driven strategies with high-level multi-reference methods might hold the solution to the challenges evidenced in this work for Dy molecules.

In conclusion, we have provided the first systematic investigation of the role of electronic correlation beyond the active space in the prediction of spin relaxation of mononuclear coordination compounds of Co(II) and Dy(III). Our results provide important evidence that quantitative predictions can be achieved for Co-based molecules by employing CASPT2, while they suggest that further development is necessary for Dy compounds. We anticipate that the long-sought goal of quantitative predictions of spin-phonon relaxation times is finally within reach and could be achieved through further investment into the development and benchmarking of multireference electronic structure methods.

\vspace{0.2cm}
\noindent
\section{Acknowledgments}
This work was funded by the Division of Chemical Sciences, Geosciences, and Biosciences, Office of Basic Energy Sciences, U.S. Department of Energy, through Grant DE-SC002183. This work has also received funding from the European Research Council (ERC) under the European Union’s Horizon 2020 research and innovation programme (grant agreement No. [948493]). Computational resources were provided by the Research Computing Center (RCC) at the University of Chicago, the Trinity College Research IT and the Irish Centre for High-End Computing (ICHEC).

\begin{suppinfo}

Parity plots comparing the numerical derivatives of crystal field parameters for complex 1 and 2; absolute energies of the lowest Kramers doublets for complex 3, total spin-phonon relaxation time computed by different methods at different temperatures for complex 1, 2, and 3, enlarged portion of plots for temperature dependence of spin-relaxation time for complex 1 and 2, sample input for MolForge software.

\end{suppinfo}

\providecommand{\latin}[1]{#1}
\makeatletter
\providecommand{\doi}
  {\begingroup\let\do\@makeother\dospecials
  \catcode`\{=1 \catcode`\}=2 \doi@aux}
\providecommand{\doi@aux}[1]{\endgroup\texttt{#1}}
\makeatother
\providecommand*\mcitethebibliography{\thebibliography}
\csname @ifundefined\endcsname{endmcitethebibliography}
  {\let\endmcitethebibliography\endthebibliography}{}


\begin{mcitethebibliography}{69}
\providecommand*\natexlab[1]{#1}
\providecommand*\mciteSetBstSublistMode[1]{}
\providecommand*\mciteSetBstMaxWidthForm[2]{}
\providecommand*\mciteBstWouldAddEndPuncttrue
  {\def\EndOfBibitem{\unskip.}}
\providecommand*\mciteBstWouldAddEndPunctfalse
  {\let\EndOfBibitem\relax}
\providecommand*\mciteSetBstMidEndSepPunct[3]{}
\providecommand*\mciteSetBstSublistLabelBeginEnd[3]{}
\providecommand*\EndOfBibitem{}
\mciteSetBstSublistMode{f}
\mciteSetBstMaxWidthForm{subitem}{(\alph{mcitesubitemcount})}
\mciteSetBstSublistLabelBeginEnd
  {\mcitemaxwidthsubitemform\space}
  {\relax}
  {\relax}

\bibitem[Leuenberger and Loss(2001)Leuenberger, and
  Loss]{leuenberger2001quantum}
Leuenberger,~M.~N.; Loss,~D. Quantum computing in molecular magnets.
  \emph{Nature} \textbf{2001}, \emph{410}, 789--793\relax
\mciteBstWouldAddEndPuncttrue
\mciteSetBstMidEndSepPunct{\mcitedefaultmidpunct}
{\mcitedefaultendpunct}{\mcitedefaultseppunct}\relax
\EndOfBibitem
\bibitem[Sessoli \latin{et~al.}(1993)Sessoli, Gatteschi, Caneschi, and
  Novak]{sessoli1993magnetic}
Sessoli,~R.; Gatteschi,~D.; Caneschi,~A.; Novak,~M. Magnetic bistability in a
  metal-ion cluster. \emph{Nature} \textbf{1993}, \emph{365}, 141--143\relax
\mciteBstWouldAddEndPuncttrue
\mciteSetBstMidEndSepPunct{\mcitedefaultmidpunct}
{\mcitedefaultendpunct}{\mcitedefaultseppunct}\relax
\EndOfBibitem
\bibitem[Rocha \latin{et~al.}(2006)Rocha, Garc\'{\i}a-Su\'arez, Bailey,
  Lambert, Ferrer, and Sanvito]{PhysRevB.73.085414}
Rocha,~A.~R.; Garc\'{\i}a-Su\'arez,~V.~M.; Bailey,~S.; Lambert,~C.; Ferrer,~J.;
  Sanvito,~S. Spin and molecular electronics in atomically generated orbital
  landscapes. \emph{Phys. Rev. B} \textbf{2006}, \emph{73}, 085414\relax
\mciteBstWouldAddEndPuncttrue
\mciteSetBstMidEndSepPunct{\mcitedefaultmidpunct}
{\mcitedefaultendpunct}{\mcitedefaultseppunct}\relax
\EndOfBibitem
\bibitem[Kragskow \latin{et~al.}(2023)Kragskow, Mattioni, Staab, Reta, Skelton,
  and Chilton]{D2CS00705C}
Kragskow,~J. G.~C.; Mattioni,~A.; Staab,~J.~K.; Reta,~D.; Skelton,~J.~M.;
  Chilton,~N.~F. Spin–phonon coupling and magnetic relaxation in
  single-molecule magnets. \emph{Chem. Soc. Rev.} \textbf{2023}, \emph{52},
  4567--4585\relax
\mciteBstWouldAddEndPuncttrue
\mciteSetBstMidEndSepPunct{\mcitedefaultmidpunct}
{\mcitedefaultendpunct}{\mcitedefaultseppunct}\relax
\EndOfBibitem
\bibitem[Lunghi(2023)]{Lunghi2023}
Lunghi,~A. In \emph{Computational Modelling of Molecular Nanomagnets};
  Rajaraman,~G., Ed.; Springer International Publishing: Cham, 2023; pp
  219--289\relax
\mciteBstWouldAddEndPuncttrue
\mciteSetBstMidEndSepPunct{\mcitedefaultmidpunct}
{\mcitedefaultendpunct}{\mcitedefaultseppunct}\relax
\EndOfBibitem
\bibitem[Lunghi \latin{et~al.}(2017)Lunghi, Totti, Sanvito, and
  Sessoli]{C7SC02832F}
Lunghi,~A.; Totti,~F.; Sanvito,~S.; Sessoli,~R. Intra-molecular origin of the
  spin-phonon coupling in slow-relaxing molecular magnets. \emph{Chem. Sci.}
  \textbf{2017}, \emph{8}, 6051--6059\relax
\mciteBstWouldAddEndPuncttrue
\mciteSetBstMidEndSepPunct{\mcitedefaultmidpunct}
{\mcitedefaultendpunct}{\mcitedefaultseppunct}\relax
\EndOfBibitem
\bibitem[Briganti \latin{et~al.}(2021)Briganti, Santanni, Tesi, Totti, Sessoli,
  and Lunghi]{doi:10.1021/jacs.1c05068}
Briganti,~M.; Santanni,~F.; Tesi,~L.; Totti,~F.; Sessoli,~R.; Lunghi,~A. A
  Complete Ab Initio View of Orbach and Raman Spin–Lattice Relaxation in a
  Dysprosium Coordination Compound. \emph{J. Am. Chem. Soc.} \textbf{2021},
  \emph{143}, 13633--13645\relax
\mciteBstWouldAddEndPuncttrue
\mciteSetBstMidEndSepPunct{\mcitedefaultmidpunct}
{\mcitedefaultendpunct}{\mcitedefaultseppunct}\relax
\EndOfBibitem
\bibitem[Mondal and Lunghi(2022)Mondal, and Lunghi]{doi:10.1021/jacs.2c08876}
Mondal,~S.; Lunghi,~A. Unraveling the Contributions to Spin–Lattice
  Relaxation in Kramers Single-Molecule Magnets. \emph{J. Am. Chem. Soc.}
  \textbf{2022}, \emph{144}, 22965--22975\relax
\mciteBstWouldAddEndPuncttrue
\mciteSetBstMidEndSepPunct{\mcitedefaultmidpunct}
{\mcitedefaultendpunct}{\mcitedefaultseppunct}\relax
\EndOfBibitem
\bibitem[Lunghi \latin{et~al.}(2017)Lunghi, Totti, Sessoli, and
  Sanvito]{lunghi2017role}
Lunghi,~A.; Totti,~F.; Sessoli,~R.; Sanvito,~S. The role of anharmonic phonons
  in under-barrier spin relaxation of single molecule magnets. \emph{Nat.
  Commun.} \textbf{2017}, \emph{8}, 14620\relax
\mciteBstWouldAddEndPuncttrue
\mciteSetBstMidEndSepPunct{\mcitedefaultmidpunct}
{\mcitedefaultendpunct}{\mcitedefaultseppunct}\relax
\EndOfBibitem
\bibitem[Escalera-Moreno \latin{et~al.}(2017)Escalera-Moreno, Suaud,
  Gaita-Ariño, and Coronado]{doi:10.1021/acs.jpclett.7b00479}
Escalera-Moreno,~L.; Suaud,~N.; Gaita-Ariño,~A.; Coronado,~E. Determining Key
  Local Vibrations in the Relaxation of Molecular Spin Qubits and
  Single-Molecule Magnets. \emph{J. Phys. Chem. Lett.} \textbf{2017}, \emph{8},
  1695--1700\relax
\mciteBstWouldAddEndPuncttrue
\mciteSetBstMidEndSepPunct{\mcitedefaultmidpunct}
{\mcitedefaultendpunct}{\mcitedefaultseppunct}\relax
\EndOfBibitem
\bibitem[Lunghi and Sanvito(2020)Lunghi, and Sanvito]{10.1063/5.0017118}
Lunghi,~A.; Sanvito,~S. {Multiple spin–phonon relaxation pathways in a Kramer
  single-ion magnet}. \emph{J. Chem. Phys.} \textbf{2020}, \emph{153},
  174113\relax
\mciteBstWouldAddEndPuncttrue
\mciteSetBstMidEndSepPunct{\mcitedefaultmidpunct}
{\mcitedefaultendpunct}{\mcitedefaultseppunct}\relax
\EndOfBibitem
\bibitem[Reta \latin{et~al.}(2021)Reta, Kragskow, and
  Chilton]{doi:10.1021/jacs.1c01410}
Reta,~D.; Kragskow,~J. G.~C.; Chilton,~N.~F. Ab Initio Prediction of
  High-Temperature Magnetic Relaxation Rates in Single-Molecule Magnets.
  \emph{J. Am. Chem. Soc.} \textbf{2021}, \emph{143}, 5943--5950\relax
\mciteBstWouldAddEndPuncttrue
\mciteSetBstMidEndSepPunct{\mcitedefaultmidpunct}
{\mcitedefaultendpunct}{\mcitedefaultseppunct}\relax
\EndOfBibitem
\bibitem[Lunghi(2022)]{doi:10.1126/sciadv.abn7880}
Lunghi,~A. Toward exact predictions of spin-phonon relaxation times: An ab
  initio implementation of open quantum systems theory. \emph{Sci. Adv.}
  \textbf{2022}, \emph{8}, eabn7880\relax
\mciteBstWouldAddEndPuncttrue
\mciteSetBstMidEndSepPunct{\mcitedefaultmidpunct}
{\mcitedefaultendpunct}{\mcitedefaultseppunct}\relax
\EndOfBibitem
\bibitem[Nabi \latin{et~al.}(2024)Nabi, Atkinson, Staab, Skelton, and
  Chilton]{D4CC03768E}
Nabi,~R.; Atkinson,~B.~E.; Staab,~J.~K.; Skelton,~J.~M.; Chilton,~N.~F. The
  impact of low-energy phonon lifetimes on the magnetic relaxation in a
  dysprosocenium single-molecule magnet. \emph{Chem. Commun.} \textbf{2024},
  \emph{60}, 13915--13918\relax
\mciteBstWouldAddEndPuncttrue
\mciteSetBstMidEndSepPunct{\mcitedefaultmidpunct}
{\mcitedefaultendpunct}{\mcitedefaultseppunct}\relax
\EndOfBibitem
\bibitem[Roos \latin{et~al.}(1980)Roos, Taylor, and Sigbahn]{roos1980complete}
Roos,~B.~O.; Taylor,~P.~R.; Sigbahn,~P.~E. A complete active space SCF method
  (CASSCF) using a density matrix formulated super-CI approach. \emph{Chem.
  Phys.} \textbf{1980}, \emph{48}, 157--173\relax
\mciteBstWouldAddEndPuncttrue
\mciteSetBstMidEndSepPunct{\mcitedefaultmidpunct}
{\mcitedefaultendpunct}{\mcitedefaultseppunct}\relax
\EndOfBibitem
\bibitem[Siegbahn \latin{et~al.}(1981)Siegbahn, Alml{\"o}f, Heiberg, and
  Roos]{siegbahn1981complete}
Siegbahn,~P.~E.; Alml{\"o}f,~J.; Heiberg,~A.; Roos,~B.~O. The complete active
  space SCF (CASSCF) method in a Newton--Raphson formulation with application
  to the HNO molecule. \emph{J. Chem. Phys.} \textbf{1981}, \emph{74},
  2384--2396\relax
\mciteBstWouldAddEndPuncttrue
\mciteSetBstMidEndSepPunct{\mcitedefaultmidpunct}
{\mcitedefaultendpunct}{\mcitedefaultseppunct}\relax
\EndOfBibitem
\bibitem[Siegbahn \latin{et~al.}(1980)Siegbahn, Heiberg, Roos, and
  Levy]{siegbahn1980comparison}
Siegbahn,~P.; Heiberg,~A.; Roos,~B.; Levy,~B. A comparison of the super-CI and
  the Newton-Raphson scheme in the complete active space SCF method.
  \emph{Phys. Scr.} \textbf{1980}, \emph{21}, 323\relax
\mciteBstWouldAddEndPuncttrue
\mciteSetBstMidEndSepPunct{\mcitedefaultmidpunct}
{\mcitedefaultendpunct}{\mcitedefaultseppunct}\relax
\EndOfBibitem
\bibitem[Angeli \latin{et~al.}(2002)Angeli, Evangelisti, Cimiraglia, and
  Maynau]{10.1063/1.1521434}
Angeli,~C.; Evangelisti,~S.; Cimiraglia,~R.; Maynau,~D. A novel
  perturbation-based complete active space–self-consistent-field algorithm:
  Application to the direct calculation of localized orbitals. \emph{J. Chem.
  Phys.} \textbf{2002}, \emph{117}, 10525--10533\relax
\mciteBstWouldAddEndPuncttrue
\mciteSetBstMidEndSepPunct{\mcitedefaultmidpunct}
{\mcitedefaultendpunct}{\mcitedefaultseppunct}\relax
\EndOfBibitem
\bibitem[Angeli \latin{et~al.}(2001)Angeli, Cimiraglia, Evangelisti, Leininger,
  and Malrieu]{10.1063/1.1361246}
Angeli,~C.; Cimiraglia,~R.; Evangelisti,~S.; Leininger,~T.; Malrieu,~J.-P.
  Introduction of n-electron valence states for multireference perturbation
  theory. \emph{J. Chem. Phys.} \textbf{2001}, \emph{114}, 10252--10264\relax
\mciteBstWouldAddEndPuncttrue
\mciteSetBstMidEndSepPunct{\mcitedefaultmidpunct}
{\mcitedefaultendpunct}{\mcitedefaultseppunct}\relax
\EndOfBibitem
\bibitem[Angeli \latin{et~al.}(2001)Angeli, Cimiraglia, and
  Malrieu]{ANGELI2001297}
Angeli,~C.; Cimiraglia,~R.; Malrieu,~J.-P. N-electron valence state
  perturbation theory: a fast implementation of the strongly contracted
  variant. \emph{Chem. Phys. Lett.} \textbf{2001}, \emph{350}, 297--305\relax
\mciteBstWouldAddEndPuncttrue
\mciteSetBstMidEndSepPunct{\mcitedefaultmidpunct}
{\mcitedefaultendpunct}{\mcitedefaultseppunct}\relax
\EndOfBibitem
\bibitem[Andersson \latin{et~al.}(1990)Andersson, Malmqvist, Roos, Sadlej, and
  Wolinski]{doi:10.1021/j100377a012}
Andersson,~K.; Malmqvist,~P.~A.; Roos,~B.~O.; Sadlej,~A.~J.; Wolinski,~K.
  Second-order perturbation theory with a CASSCF reference function. \emph{J.
  Phys. Chem.} \textbf{1990}, \emph{94}, 5483--5488\relax
\mciteBstWouldAddEndPuncttrue
\mciteSetBstMidEndSepPunct{\mcitedefaultmidpunct}
{\mcitedefaultendpunct}{\mcitedefaultseppunct}\relax
\EndOfBibitem
\bibitem[Roos \latin{et~al.}(1996)Roos, Andersson, Fülscher, Serrano-Andrés,
  Pierloot, Merchán, and Molina]{ROOS1996257}
Roos,~B.~O.; Andersson,~K.; Fülscher,~M.~P.; Serrano-Andrés,~L.;
  Pierloot,~K.; Merchán,~M.; Molina,~V. Applications of level shift corrected
  perturbation theory in electronic spectroscopy. \emph{J. Mol. Struct.
  THEOCHEM} \textbf{1996}, \emph{388}, 257--276\relax
\mciteBstWouldAddEndPuncttrue
\mciteSetBstMidEndSepPunct{\mcitedefaultmidpunct}
{\mcitedefaultendpunct}{\mcitedefaultseppunct}\relax
\EndOfBibitem
\bibitem[Li~Manni \latin{et~al.}(2014)Li~Manni, Carlson, Luo, Ma, Olsen,
  Truhlar, and Gagliardi]{doi:10.1021/ct500483t}
Li~Manni,~G.; Carlson,~R.~K.; Luo,~S.; Ma,~D.; Olsen,~J.; Truhlar,~D.~G.;
  Gagliardi,~L. Multiconfiguration Pair-Density Functional Theory. \emph{J.
  Chem. Theory Comput.} \textbf{2014}, \emph{10}, 3669--3680\relax
\mciteBstWouldAddEndPuncttrue
\mciteSetBstMidEndSepPunct{\mcitedefaultmidpunct}
{\mcitedefaultendpunct}{\mcitedefaultseppunct}\relax
\EndOfBibitem
\bibitem[Zhou \latin{et~al.}(2022)Zhou, Hermes, Wu, Bao, Pandharkar, King,
  Zhang, Scott, Lykhin, Gagliardi, and Truhlar]{D2SC01022D}
Zhou,~C.; Hermes,~M.~R.; Wu,~D.; Bao,~J.~J.; Pandharkar,~R.; King,~D.~S.;
  Zhang,~D.; Scott,~T.~R.; Lykhin,~A.~O.; Gagliardi,~L.; Truhlar,~D.~G.
  Electronic structure of strongly correlated systems: recent developments in
  multiconfiguration pair-density functional theory and multiconfiguration
  nonclassical-energy functional theory. \emph{Chem. Sci.} \textbf{2022},
  \emph{13}, 7685--7706\relax
\mciteBstWouldAddEndPuncttrue
\mciteSetBstMidEndSepPunct{\mcitedefaultmidpunct}
{\mcitedefaultendpunct}{\mcitedefaultseppunct}\relax
\EndOfBibitem
\bibitem[Sharma \latin{et~al.}(2021)Sharma, Bao, Truhlar, and
  Gagliardi]{annurev:/content/journals/10.1146/annurev-physchem-090419-043839}
Sharma,~P.; Bao,~J.~J.; Truhlar,~D.~G.; Gagliardi,~L. Multiconfiguration
  Pair-Density Functional Theory. \emph{Annu. Rev. Phys. Chem.} \textbf{2021},
  \emph{72}, 541--564\relax
\mciteBstWouldAddEndPuncttrue
\mciteSetBstMidEndSepPunct{\mcitedefaultmidpunct}
{\mcitedefaultendpunct}{\mcitedefaultseppunct}\relax
\EndOfBibitem
\bibitem[Gagliardi \latin{et~al.}(2017)Gagliardi, Truhlar, Li~Manni, Carlson,
  Hoyer, and Bao]{doi:10.1021/acs.accounts.6b00471}
Gagliardi,~L.; Truhlar,~D.~G.; Li~Manni,~G.; Carlson,~R.~K.; Hoyer,~C.~E.;
  Bao,~J.~L. Multiconfiguration Pair-Density Functional Theory: A New Way To
  Treat Strongly Correlated Systems. \emph{Acc. Chem. Res.} \textbf{2017},
  \emph{50}, 66--73\relax
\mciteBstWouldAddEndPuncttrue
\mciteSetBstMidEndSepPunct{\mcitedefaultmidpunct}
{\mcitedefaultendpunct}{\mcitedefaultseppunct}\relax
\EndOfBibitem
\bibitem[Ungur and Chibotaru(2017)Ungur, and
  Chibotaru]{https://doi.org/10.1002/chem.201605102}
Ungur,~L.; Chibotaru,~L.~F. Ab Initio Crystal Field for Lanthanides.
  \emph{Chem. Eur. J.} \textbf{2017}, \emph{23}, 3708--3718\relax
\mciteBstWouldAddEndPuncttrue
\mciteSetBstMidEndSepPunct{\mcitedefaultmidpunct}
{\mcitedefaultendpunct}{\mcitedefaultseppunct}\relax
\EndOfBibitem
\bibitem[Weller \latin{et~al.}(2023)Weller, Atanasov, Demeshko, Chen, Mohelsky,
  Bill, Orlita, Meyer, Neese, and Werncke]{doi:10.1021/acs.inorgchem.2c04050}
Weller,~R.; Atanasov,~M.; Demeshko,~S.; Chen,~T.-Y.; Mohelsky,~I.; Bill,~E.;
  Orlita,~M.; Meyer,~F.; Neese,~F.; Werncke,~C.~G. On the Single-Molecule
  Magnetic Behavior of Linear Iron(I) Arylsilylamides. \emph{Inorg. Chem.}
  \textbf{2023}, \emph{62}, 3153--3161\relax
\mciteBstWouldAddEndPuncttrue
\mciteSetBstMidEndSepPunct{\mcitedefaultmidpunct}
{\mcitedefaultendpunct}{\mcitedefaultseppunct}\relax
\EndOfBibitem
\bibitem[Atanasov \latin{et~al.}(2022)Atanasov, Spiller, and
  Neese]{doi:10.1039/D2CP02975H}
Atanasov,~M.; Spiller,~N.; Neese,~F. Magnetic exchange and valence
  delocalization in a mixed valence [Fe\textsuperscript{2+}
  Fe\textsuperscript{3+} Te\textsubscript{2}]\textsuperscript{+} complex:
  insights from theory and interpretations of magnetic and spectroscopic data.
  \emph{Phys. Chem. Chem. Phys.} \textbf{2022}, \emph{24}, 20760--20775\relax
\mciteBstWouldAddEndPuncttrue
\mciteSetBstMidEndSepPunct{\mcitedefaultmidpunct}
{\mcitedefaultendpunct}{\mcitedefaultseppunct}\relax
\EndOfBibitem
\bibitem[Suturina \latin{et~al.}(2015)Suturina, Maganas, Bill, Atanasov, and
  Neese]{doi:10.1021/acs.inorgchem.5b01706}
Suturina,~E.~A.; Maganas,~D.; Bill,~E.; Atanasov,~M.; Neese,~F.
  Magneto-Structural Correlations in a Series of Pseudotetrahedral
  [Co\textsuperscript{II}(XR)\textsubscript{4}]\textsuperscript{2–} Single
  Molecule Magnets: An ab Initio Ligand Field Study. \emph{Inorg. Chem.}
  \textbf{2015}, \emph{54}, 9948--9961\relax
\mciteBstWouldAddEndPuncttrue
\mciteSetBstMidEndSepPunct{\mcitedefaultmidpunct}
{\mcitedefaultendpunct}{\mcitedefaultseppunct}\relax
\EndOfBibitem
\bibitem[Atanasov and Neese(2018)Atanasov, and Neese]{Atanasov_2018}
Atanasov,~M.; Neese,~F. Computational Studies on Vibronic Coupling in Single
  Molecule Magnets: Impact on the Mechanism of Magnetic Relaxation. \emph{J.
  Phys. Conf. Ser.} \textbf{2018}, \emph{1148}, 012006\relax
\mciteBstWouldAddEndPuncttrue
\mciteSetBstMidEndSepPunct{\mcitedefaultmidpunct}
{\mcitedefaultendpunct}{\mcitedefaultseppunct}\relax
\EndOfBibitem
\bibitem[Atanasov \latin{et~al.}(2012)Atanasov, Ganyushin, Sivalingam, and
  Neese]{atanasov2012modern}
Atanasov,~M.; Ganyushin,~D.; Sivalingam,~K.; Neese,~F. A modern
  first-principles view on ligand field theory through the eyes of correlated
  multireference wavefunctions. \emph{Molecular electronic structures of
  transition metal complexes II} \textbf{2012}, 149--220\relax
\mciteBstWouldAddEndPuncttrue
\mciteSetBstMidEndSepPunct{\mcitedefaultmidpunct}
{\mcitedefaultendpunct}{\mcitedefaultseppunct}\relax
\EndOfBibitem
\bibitem[Singh \latin{et~al.}(2018)Singh, Atanasov, and
  Neese]{singh2018challenges}
Singh,~S.~K.; Atanasov,~M.; Neese,~F. Challenges in multireference perturbation
  theory for the calculations of the g-tensor of first-row transition-metal
  complexes. \emph{J. Chem. Theory Comput.} \textbf{2018}, \emph{14},
  4662--4677\relax
\mciteBstWouldAddEndPuncttrue
\mciteSetBstMidEndSepPunct{\mcitedefaultmidpunct}
{\mcitedefaultendpunct}{\mcitedefaultseppunct}\relax
\EndOfBibitem
\bibitem[Aravena \latin{et~al.}(2016)Aravena, Atanasov, and
  Neese]{doi:10.1021/acs.inorgchem.6b00244}
Aravena,~D.; Atanasov,~M.; Neese,~F. Periodic Trends in Lanthanide Compounds
  through the Eyes of Multireference ab Initio Theory. \emph{Inorg. Chem.}
  \textbf{2016}, \emph{55}, 4457--4469\relax
\mciteBstWouldAddEndPuncttrue
\mciteSetBstMidEndSepPunct{\mcitedefaultmidpunct}
{\mcitedefaultendpunct}{\mcitedefaultseppunct}\relax
\EndOfBibitem
\bibitem[Jung \latin{et~al.}(2017)Jung, Atanasov, and Neese]{jung2017ab}
Jung,~J.; Atanasov,~M.; Neese,~F. Ab initio ligand-field theory analysis and
  covalency trends in actinide and lanthanide free ions and octahedral
  complexes. \emph{Inorg. Chem.} \textbf{2017}, \emph{56}, 8802--8816\relax
\mciteBstWouldAddEndPuncttrue
\mciteSetBstMidEndSepPunct{\mcitedefaultmidpunct}
{\mcitedefaultendpunct}{\mcitedefaultseppunct}\relax
\EndOfBibitem
\bibitem[Fataftah \latin{et~al.}(2014)Fataftah, Zadrozny, Rogers, and
  Freedman]{doi:10.1021/ic501906z}
Fataftah,~M.~S.; Zadrozny,~J.~M.; Rogers,~D.~M.; Freedman,~D.~E. A Mononuclear
  Transition Metal Single-Molecule Magnet in a Nuclear Spin-Free Ligand
  Environment. \emph{Inorg. Chem.} \textbf{2014}, \emph{53}, 10716--10721\relax
\mciteBstWouldAddEndPuncttrue
\mciteSetBstMidEndSepPunct{\mcitedefaultmidpunct}
{\mcitedefaultendpunct}{\mcitedefaultseppunct}\relax
\EndOfBibitem
\bibitem[Rechkemmer \latin{et~al.}(2016)Rechkemmer, Breitgoff, Van Der~Meer,
  Atanasov, Hakl, Orlita, Neugebauer, Neese, Sarkar, and
  Van~Slageren]{rechkemmer2016four}
Rechkemmer,~Y.; Breitgoff,~F.~D.; Van Der~Meer,~M.; Atanasov,~M.; Hakl,~M.;
  Orlita,~M.; Neugebauer,~P.; Neese,~F.; Sarkar,~B.; Van~Slageren,~J. A
  four-coordinate cobalt (II) single-ion magnet with coercivity and a very high
  energy barrier. \emph{Nat. Commun.} \textbf{2016}, \emph{7}, 10467\relax
\mciteBstWouldAddEndPuncttrue
\mciteSetBstMidEndSepPunct{\mcitedefaultmidpunct}
{\mcitedefaultendpunct}{\mcitedefaultseppunct}\relax
\EndOfBibitem
\bibitem[Liu \latin{et~al.}(2016)Liu, Chen, Liu, Vieru, Ungur, Jia, Chibotaru,
  Lan, Wernsdorfer, Gao, Chen, and Tong]{doi:10.1021/jacs.6b02638}
Liu,~J.; Chen,~Y.-C.; Liu,~J.-L.; Vieru,~V.; Ungur,~L.; Jia,~J.-H.;
  Chibotaru,~L.~F.; Lan,~Y.; Wernsdorfer,~W.; Gao,~S.; Chen,~X.-M.; Tong,~M.-L.
  A Stable Pentagonal Bipyramidal Dy(III) Single-Ion Magnet with a Record
  Magnetization Reversal Barrier over 1000 K. \emph{J. Am. Chem. Soc.}
  \textbf{2016}, \emph{138}, 5441--5450\relax
\mciteBstWouldAddEndPuncttrue
\mciteSetBstMidEndSepPunct{\mcitedefaultmidpunct}
{\mcitedefaultendpunct}{\mcitedefaultseppunct}\relax
\EndOfBibitem
\bibitem[Roos \latin{et~al.}(1996)Roos, Andersson, Fulscher, Malmqvist,
  Serrano-Andres, Pierloot, and
  Merchan]{doi:https://doi.org/10.1002/9780470141526.ch5}
Roos,~B.~O.; Andersson,~K.; Fulscher,~M.~P.; Malmqvist,~P.-a.;
  Serrano-Andres,~L.; Pierloot,~K.; Merchan,~M. \emph{Advances in Chemical
  Physics}; John Wiley \& Sons, Ltd, 1996; Vol. XCIII; pp 219--331\relax
\mciteBstWouldAddEndPuncttrue
\mciteSetBstMidEndSepPunct{\mcitedefaultmidpunct}
{\mcitedefaultendpunct}{\mcitedefaultseppunct}\relax
\EndOfBibitem
\bibitem[Andersson \latin{et~al.}(1992)Andersson, Malmqvist, and
  Roos]{10.1063/1.462209}
Andersson,~K.; Malmqvist,~P.; Roos,~B.~O. Second‐order perturbation theory
  with a complete active space self‐consistent field reference function.
  \emph{J. Chem. Phys.} \textbf{1992}, \emph{96}, 1218--1226\relax
\mciteBstWouldAddEndPuncttrue
\mciteSetBstMidEndSepPunct{\mcitedefaultmidpunct}
{\mcitedefaultendpunct}{\mcitedefaultseppunct}\relax
\EndOfBibitem
\bibitem[Pulay(2011)]{https://doi.org/10.1002/qua.23052}
Pulay,~P. A perspective on the CASPT2 method. \emph{Int. J. Quantum Chem.}
  \textbf{2011}, \emph{111}, 3273--3279\relax
\mciteBstWouldAddEndPuncttrue
\mciteSetBstMidEndSepPunct{\mcitedefaultmidpunct}
{\mcitedefaultendpunct}{\mcitedefaultseppunct}\relax
\EndOfBibitem
\bibitem[Kats and Werner(2019)Kats, and Werner]{10.1063/1.5097644}
Kats,~D.; Werner,~H.-J. Multi-state local complete active space second-order
  perturbation theory using pair natural orbitals (PNO-MS-CASPT2). \emph{J.
  Chem. Phys.} \textbf{2019}, \emph{150}, 214107\relax
\mciteBstWouldAddEndPuncttrue
\mciteSetBstMidEndSepPunct{\mcitedefaultmidpunct}
{\mcitedefaultendpunct}{\mcitedefaultseppunct}\relax
\EndOfBibitem
\bibitem[Battaglia \latin{et~al.}(2023)Battaglia, {Fdez. Galván}, and
  Lindh]{BATTAGLIA2023135}
Battaglia,~S.; {Fdez. Galván},~I.; Lindh,~R. In \emph{Theoretical and
  Computational Photochemistry}; García-Iriepa,~C., Marazzi,~M., Eds.;
  Elsevier, 2023; pp 135--162\relax
\mciteBstWouldAddEndPuncttrue
\mciteSetBstMidEndSepPunct{\mcitedefaultmidpunct}
{\mcitedefaultendpunct}{\mcitedefaultseppunct}\relax
\EndOfBibitem
\bibitem[Perdew \latin{et~al.}(1996)Perdew, Burke, and
  Ernzerhof]{PhysRevLett.77.3865}
Perdew,~J.~P.; Burke,~K.; Ernzerhof,~M. Generalized Gradient Approximation Made
  Simple. \emph{Phys. Rev. Lett.} \textbf{1996}, \emph{77}, 3865--3868\relax
\mciteBstWouldAddEndPuncttrue
\mciteSetBstMidEndSepPunct{\mcitedefaultmidpunct}
{\mcitedefaultendpunct}{\mcitedefaultseppunct}\relax
\EndOfBibitem
\bibitem[Odoh \latin{et~al.}(2016)Odoh, Manni, Carlson, Truhlar, and
  Gagliardi]{C5SC03321G}
Odoh,~S.~O.; Manni,~G.~L.; Carlson,~R.~K.; Truhlar,~D.~G.; Gagliardi,~L.
  Separated-pair approximation and separated-pair pair-density functional
  theory. \emph{Chem. Sci.} \textbf{2016}, \emph{7}, 2399--2413\relax
\mciteBstWouldAddEndPuncttrue
\mciteSetBstMidEndSepPunct{\mcitedefaultmidpunct}
{\mcitedefaultendpunct}{\mcitedefaultseppunct}\relax
\EndOfBibitem
\bibitem[Bao \latin{et~al.}(2017)Bao, Odoh, Gagliardi, and
  Truhlar]{doi:10.1021/acs.jctc.6b01102}
Bao,~J.~L.; Odoh,~S.~O.; Gagliardi,~L.; Truhlar,~D.~G. Predicting Bond
  Dissociation Energies of Transition-Metal Compounds by Multiconfiguration
  Pair-Density Functional Theory and Second-Order Perturbation Theory Based on
  Correlated Participating Orbitals and Separated Pairs. \emph{J. Chem. Theory
  Comput.} \textbf{2017}, \emph{13}, 616--626\relax
\mciteBstWouldAddEndPuncttrue
\mciteSetBstMidEndSepPunct{\mcitedefaultmidpunct}
{\mcitedefaultendpunct}{\mcitedefaultseppunct}\relax
\EndOfBibitem
\bibitem[Stoneburner \latin{et~al.}(2018)Stoneburner, Truhlar, and
  Gagliardi]{10.1063/1.5017132}
Stoneburner,~S.~J.; Truhlar,~D.~G.; Gagliardi,~L. MC-PDFT can calculate
  singlet–triplet splittings of organic diradicals. \emph{J. Chem. Phys.}
  \textbf{2018}, \emph{148}, 064108\relax
\mciteBstWouldAddEndPuncttrue
\mciteSetBstMidEndSepPunct{\mcitedefaultmidpunct}
{\mcitedefaultendpunct}{\mcitedefaultseppunct}\relax
\EndOfBibitem
\bibitem[Ghosh \latin{et~al.}(2017)Ghosh, Cramer, Truhlar, and
  Gagliardi]{C6SC05036K}
Ghosh,~S.; Cramer,~C.~J.; Truhlar,~D.~G.; Gagliardi,~L.
  Generalized-active-space pair-density functional theory: an efficient method
  to study large{,} strongly correlated{,} conjugated systems. \emph{Chem.
  Sci.} \textbf{2017}, \emph{8}, 2741--2750\relax
\mciteBstWouldAddEndPuncttrue
\mciteSetBstMidEndSepPunct{\mcitedefaultmidpunct}
{\mcitedefaultendpunct}{\mcitedefaultseppunct}\relax
\EndOfBibitem
\bibitem[Wilbraham \latin{et~al.}(2017)Wilbraham, Verma, Truhlar, Gagliardi,
  and Ciofini]{doi:10.1021/acs.jpclett.7b00570}
Wilbraham,~L.; Verma,~P.; Truhlar,~D.~G.; Gagliardi,~L.; Ciofini,~I.
  Multiconfiguration Pair-Density Functional Theory Predicts Spin-State
  Ordering in Iron Complexes with the Same Accuracy as Complete Active Space
  Second-Order Perturbation Theory at a Significantly Reduced Computational
  Cost. \emph{J. Phys. Chem. Lett.} \textbf{2017}, \emph{8}, 2026--2030\relax
\mciteBstWouldAddEndPuncttrue
\mciteSetBstMidEndSepPunct{\mcitedefaultmidpunct}
{\mcitedefaultendpunct}{\mcitedefaultseppunct}\relax
\EndOfBibitem
\bibitem[Ghosh \latin{et~al.}(2015)Ghosh, Sonnenberger, Hoyer, Truhlar, and
  Gagliardi]{doi:10.1021/acs.jctc.5b00456}
Ghosh,~S.; Sonnenberger,~A.~L.; Hoyer,~C.~E.; Truhlar,~D.~G.; Gagliardi,~L.
  Multiconfiguration Pair-Density Functional Theory Outperforms Kohn–Sham
  Density Functional Theory and Multireference Perturbation Theory for
  Ground-State and Excited-State Charge Transfer. \emph{J. Chem. Theory
  Comput.} \textbf{2015}, \emph{11}, 3643--3649\relax
\mciteBstWouldAddEndPuncttrue
\mciteSetBstMidEndSepPunct{\mcitedefaultmidpunct}
{\mcitedefaultendpunct}{\mcitedefaultseppunct}\relax
\EndOfBibitem
\bibitem[Hoyer \latin{et~al.}(2016)Hoyer, Ghosh, Truhlar, and
  Gagliardi]{doi:10.1021/acs.jpclett.5b02773}
Hoyer,~C.~E.; Ghosh,~S.; Truhlar,~D.~G.; Gagliardi,~L. Multiconfiguration
  Pair-Density Functional Theory Is as Accurate as CASPT2 for Electronic
  Excitation. \emph{J. Phys. Chem. Lett.} \textbf{2016}, \emph{7},
  586--591\relax
\mciteBstWouldAddEndPuncttrue
\mciteSetBstMidEndSepPunct{\mcitedefaultmidpunct}
{\mcitedefaultendpunct}{\mcitedefaultseppunct}\relax
\EndOfBibitem
\bibitem[Chibotaru and Ungur(2012)Chibotaru, and Ungur]{10.1063/1.4739763}
Chibotaru,~L.~F.; Ungur,~L. Ab initio calculation of anisotropic magnetic
  properties of complexes. I. Unique definition of pseudospin Hamiltonians and
  their derivation. \emph{J. Chem. Phys.} \textbf{2012}, \emph{137},
  064112\relax
\mciteBstWouldAddEndPuncttrue
\mciteSetBstMidEndSepPunct{\mcitedefaultmidpunct}
{\mcitedefaultendpunct}{\mcitedefaultseppunct}\relax
\EndOfBibitem
\bibitem[Maurice \latin{et~al.}(2009)Maurice, Bastardis, Graaf, Suaud, Mallah,
  and Guih{\'e}ry]{doi:10.1021/ct900326e}
Maurice,~R.; Bastardis,~R.; Graaf,~C.~d.; Suaud,~N.; Mallah,~T.;
  Guih{\'e}ry,~N. Universal Theoretical Approach to Extract Anisotropic Spin
  Hamiltonians. \emph{J. Chem. Theory Comput.} \textbf{2009}, \emph{5},
  2977--2984\relax
\mciteBstWouldAddEndPuncttrue
\mciteSetBstMidEndSepPunct{\mcitedefaultmidpunct}
{\mcitedefaultendpunct}{\mcitedefaultseppunct}\relax
\EndOfBibitem
\bibitem[Mariano \latin{et~al.}(2024)Mariano, Mondal, and
  Lunghi]{doi:10.1021/acs.jctc.3c01130}
Mariano,~L.~A.; Mondal,~S.; Lunghi,~A. Spin-Vibronic Dynamics in Open-Shell
  Systems beyond the Spin Hamiltonian Formalism. \emph{J. Chem. Theory Comput.}
  \textbf{2024}, \emph{20}, 323--332\relax
\mciteBstWouldAddEndPuncttrue
\mciteSetBstMidEndSepPunct{\mcitedefaultmidpunct}
{\mcitedefaultendpunct}{\mcitedefaultseppunct}\relax
\EndOfBibitem
\bibitem[Kühne \latin{et~al.}(2020)Kühne, Iannuzzi, Del~Ben, Rybkin, Seewald,
  Stein, Laino, Khaliullin, Schütt, Schiffmann, Golze, Wilhelm, Chulkov,
  Bani-Hashemian, Weber, Borštnik, Taillefumier, Jakobovits, Lazzaro, Pabst,
  Müller, Schade, Guidon, Andermatt, Holmberg, Schenter, Hehn, Bussy,
  Belleflamme, Tabacchi, Glöß, Lass, Bethune, Mundy, Plessl, Watkins,
  VandeVondele, Krack, and Hutter]{10.1063/5.0007045}
Kühne,~T.~D. \latin{et~al.}  CP2K: An electronic structure and molecular
  dynamics software package - Quickstep: Efficient and accurate electronic
  structure calculations. \emph{J. Chem. Phys.} \textbf{2020}, \emph{152},
  194103\relax
\mciteBstWouldAddEndPuncttrue
\mciteSetBstMidEndSepPunct{\mcitedefaultmidpunct}
{\mcitedefaultendpunct}{\mcitedefaultseppunct}\relax
\EndOfBibitem
\bibitem[Grimme \latin{et~al.}(2010)Grimme, Antony, Ehrlich, and
  Krieg]{10.1063/1.3382344}
Grimme,~S.; Antony,~J.; Ehrlich,~S.; Krieg,~H. A consistent and accurate ab
  initio parametrization of density functional dispersion correction (DFT-D)
  for the 94 elements H-Pu. \emph{J. Chem. Phys.} \textbf{2010}, \emph{132},
  154104\relax
\mciteBstWouldAddEndPuncttrue
\mciteSetBstMidEndSepPunct{\mcitedefaultmidpunct}
{\mcitedefaultendpunct}{\mcitedefaultseppunct}\relax
\EndOfBibitem
\bibitem[Aquilante \latin{et~al.}(2016)Aquilante, Autschbach, Carlson,
  Chibotaru, Delcey, De~Vico, Fdez.~Galván, Ferré, Frutos, Gagliardi,
  Garavelli, Giussani, Hoyer, Li~Manni, Lischka, Ma, Malmqvist, Müller, Nenov,
  Olivucci, Pedersen, Peng, Plasser, Pritchard, Reiher, Rivalta, Schapiro,
  Segarra-Martí, Stenrup, Truhlar, Ungur, Valentini, Vancoillie, Veryazov,
  Vysotskiy, Weingart, Zapata, and Lindh]{https://doi.org/10.1002/jcc.24221}
Aquilante,~F. \latin{et~al.}  Molcas 8: New capabilities for
  multiconfigurational quantum chemical calculations across the periodic table.
  \emph{J. Comput. Chem.} \textbf{2016}, \emph{37}, 506--541\relax
\mciteBstWouldAddEndPuncttrue
\mciteSetBstMidEndSepPunct{\mcitedefaultmidpunct}
{\mcitedefaultendpunct}{\mcitedefaultseppunct}\relax
\EndOfBibitem
\bibitem[Åke Malmqvist \latin{et~al.}(2002)Åke Malmqvist, Roos, and
  Schimmelpfennig]{MALMQVIST2002230}
Åke Malmqvist,~P.; Roos,~B.~O.; Schimmelpfennig,~B. The restricted active
  space (RAS) state interaction approach with spin–orbit coupling.
  \emph{Chem. Phys. Lett.} \textbf{2002}, \emph{357}, 230--240\relax
\mciteBstWouldAddEndPuncttrue
\mciteSetBstMidEndSepPunct{\mcitedefaultmidpunct}
{\mcitedefaultendpunct}{\mcitedefaultseppunct}\relax
\EndOfBibitem
\bibitem[Finley \latin{et~al.}(1998)Finley, Åke Malmqvist, Roos, and
  Serrano-Andrés]{FINLEY1998299}
Finley,~J.; Åke Malmqvist,~P.; Roos,~B.~O.; Serrano-Andrés,~L. The
  multi-state CASPT2 method. \emph{Chem. Phys. Lett.} \textbf{1998},
  \emph{288}, 299--306\relax
\mciteBstWouldAddEndPuncttrue
\mciteSetBstMidEndSepPunct{\mcitedefaultmidpunct}
{\mcitedefaultendpunct}{\mcitedefaultseppunct}\relax
\EndOfBibitem
\bibitem[Granovsky(2011)]{10.1063/1.3596699}
Granovsky,~A.~A. {Extended multi-configuration quasi-degenerate perturbation
  theory: The new approach to multi-state multi-reference perturbation theory}.
  \emph{J. Chem. Phys.} \textbf{2011}, \emph{134}, 214113\relax
\mciteBstWouldAddEndPuncttrue
\mciteSetBstMidEndSepPunct{\mcitedefaultmidpunct}
{\mcitedefaultendpunct}{\mcitedefaultseppunct}\relax
\EndOfBibitem
\bibitem[Shiozaki \latin{et~al.}(2011)Shiozaki, Győrffy, Celani, and
  Werner]{10.1063/1.3633329}
Shiozaki,~T.; Győrffy,~W.; Celani,~P.; Werner,~H.-J. {Communication: Extended
  multi-state complete active space second-order perturbation theory: Energy
  and nuclear gradients}. \emph{J. Chem. Phys.} \textbf{2011}, \emph{135},
  081106\relax
\mciteBstWouldAddEndPuncttrue
\mciteSetBstMidEndSepPunct{\mcitedefaultmidpunct}
{\mcitedefaultendpunct}{\mcitedefaultseppunct}\relax
\EndOfBibitem
\bibitem[Bao \latin{et~al.}(2020)Bao, Zhou, and
  Truhlar]{doi:10.1021/acs.jctc.0c00908}
Bao,~J.~J.; Zhou,~C.; Truhlar,~D.~G. Compressed-State Multistate Pair-Density
  Functional Theory. \emph{J. Chem. Theory Comput.} \textbf{2020}, \emph{16},
  7444--7452\relax
\mciteBstWouldAddEndPuncttrue
\mciteSetBstMidEndSepPunct{\mcitedefaultmidpunct}
{\mcitedefaultendpunct}{\mcitedefaultseppunct}\relax
\EndOfBibitem
\bibitem[Sarkar and Gagliardi(2023)Sarkar, and
  Gagliardi]{doi:10.1021/acs.jpca.3c05803}
Sarkar,~A.; Gagliardi,~L. Multiconfiguration Pair-Density Functional Theory for
  Vertical Excitation Energies in Actinide Molecules. \emph{J. Phys. Chem. A}
  \textbf{2023}, \emph{127}, 9389--9397\relax
\mciteBstWouldAddEndPuncttrue
\mciteSetBstMidEndSepPunct{\mcitedefaultmidpunct}
{\mcitedefaultendpunct}{\mcitedefaultseppunct}\relax
\EndOfBibitem
\bibitem[Hennefarth \latin{et~al.}(2023)Hennefarth, Hermes, Truhlar, and
  Gagliardi]{doi:10.1021/acs.jctc.3c00207}
Hennefarth,~M.~R.; Hermes,~M.~R.; Truhlar,~D.~G.; Gagliardi,~L. Linearized
  Pair-Density Functional Theory. \emph{J. Chem. Theory Comput.} \textbf{2023},
  \emph{19}, 3172--3183\relax
\mciteBstWouldAddEndPuncttrue
\mciteSetBstMidEndSepPunct{\mcitedefaultmidpunct}
{\mcitedefaultendpunct}{\mcitedefaultseppunct}\relax
\EndOfBibitem
\bibitem[Nabi \latin{et~al.}(2023)Nabi, Staab, Mattioni, Kragskow, Reta,
  Skelton, and Chilton]{doi:10.1021/jacs.3c06015}
Nabi,~R.; Staab,~J.~K.; Mattioni,~A.; Kragskow,~J. G.~C.; Reta,~D.;
  Skelton,~J.~M.; Chilton,~N.~F. Accurate and Efficient Spin–Phonon Coupling
  and Spin Dynamics Calculations for Molecular Solids. \emph{J. Am. Chem. Soc.}
  \textbf{2023}, \emph{145}, 24558--24567\relax
\mciteBstWouldAddEndPuncttrue
\mciteSetBstMidEndSepPunct{\mcitedefaultmidpunct}
{\mcitedefaultendpunct}{\mcitedefaultseppunct}\relax
\EndOfBibitem
\bibitem[Staab and Chilton(2022)Staab, and
  Chilton]{doi:10.1021/acs.jctc.2c00611}
Staab,~J.~K.; Chilton,~N.~F. Analytic Linear Vibronic Coupling Method for
  First-Principles Spin-Dynamics Calculations in Single-Molecule Magnets.
  \emph{J. Chem. Theory Comput.} \textbf{2022}, \emph{18}, 6588--6599, PMID:
  36269220\relax
\mciteBstWouldAddEndPuncttrue
\mciteSetBstMidEndSepPunct{\mcitedefaultmidpunct}
{\mcitedefaultendpunct}{\mcitedefaultseppunct}\relax
\EndOfBibitem
\bibitem[Lunghi and Sanvito(2022)Lunghi, and Sanvito]{lunghi2022computational}
Lunghi,~A.; Sanvito,~S. Computational design of magnetic molecules and their
  environment using quantum chemistry, machine learning and multiscale
  simulations. \emph{Nat. Rev. Chem.} \textbf{2022}, \emph{6}, 761--781\relax
\mciteBstWouldAddEndPuncttrue
\mciteSetBstMidEndSepPunct{\mcitedefaultmidpunct}
{\mcitedefaultendpunct}{\mcitedefaultseppunct}\relax
\EndOfBibitem
\bibitem[Briganti and Lunghi(2024)Briganti, and
  Lunghi]{briganti2024machinelearningframeworkacceleratingspinlattice}
Briganti,~V.; Lunghi,~A. A machine-learning framework for accelerating
  spin-lattice relaxation simulations. 2024;
  \url{https://arxiv.org/abs/2410.08912}\relax
\mciteBstWouldAddEndPuncttrue
\mciteSetBstMidEndSepPunct{\mcitedefaultmidpunct}
{\mcitedefaultendpunct}{\mcitedefaultseppunct}\relax
\EndOfBibitem
\end{mcitethebibliography}

\end{document}


\newpage
\tableofcontents

\newpage

\section{Parity plots and total spin-relaxation time at different temperatures for [Co(C\textsubscript{3}S\textsubscript{5})\textsubscript{2}](Ph\textsubscript{4}P)\textsubscript{2} (complex \textbf{1})}

\graphicspath{{Figures/}}
\begin{figure}[h!]
\centerline{\includegraphics[width=1.15\textwidth,,height=0.48\textwidth]{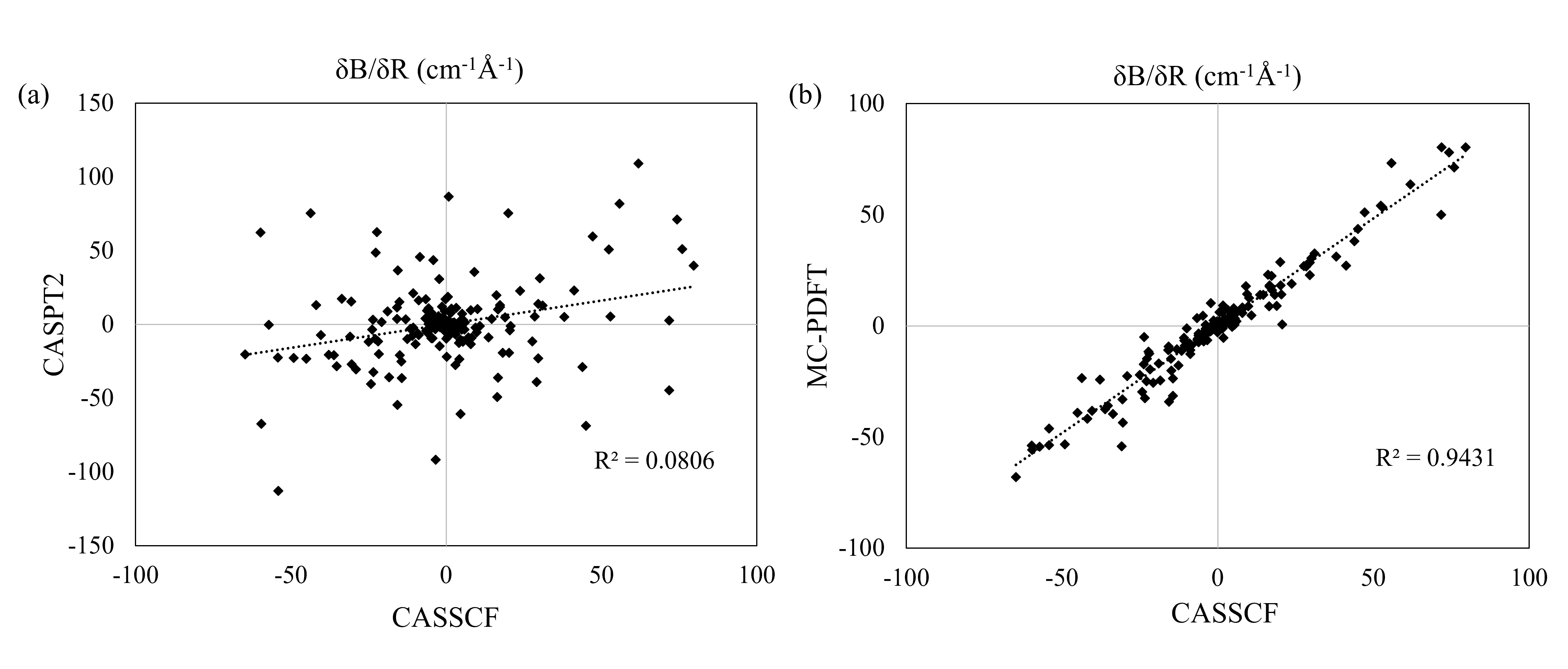}}
\centering
\caption{Parity plots comparing the numerical derivatives of the crystal field parameters computed at (a) CASSCF and CASPT2, and (b) CASSCF and MC-PDFT levels for compound \textbf{1}}
\label{fig: Parity plot mol_1}
\end{figure}

\graphicspath{{Figures/}}
\begin{figure}[h!]
\centerline{\includegraphics[width=0.62\textwidth,,height=0.48\textwidth]{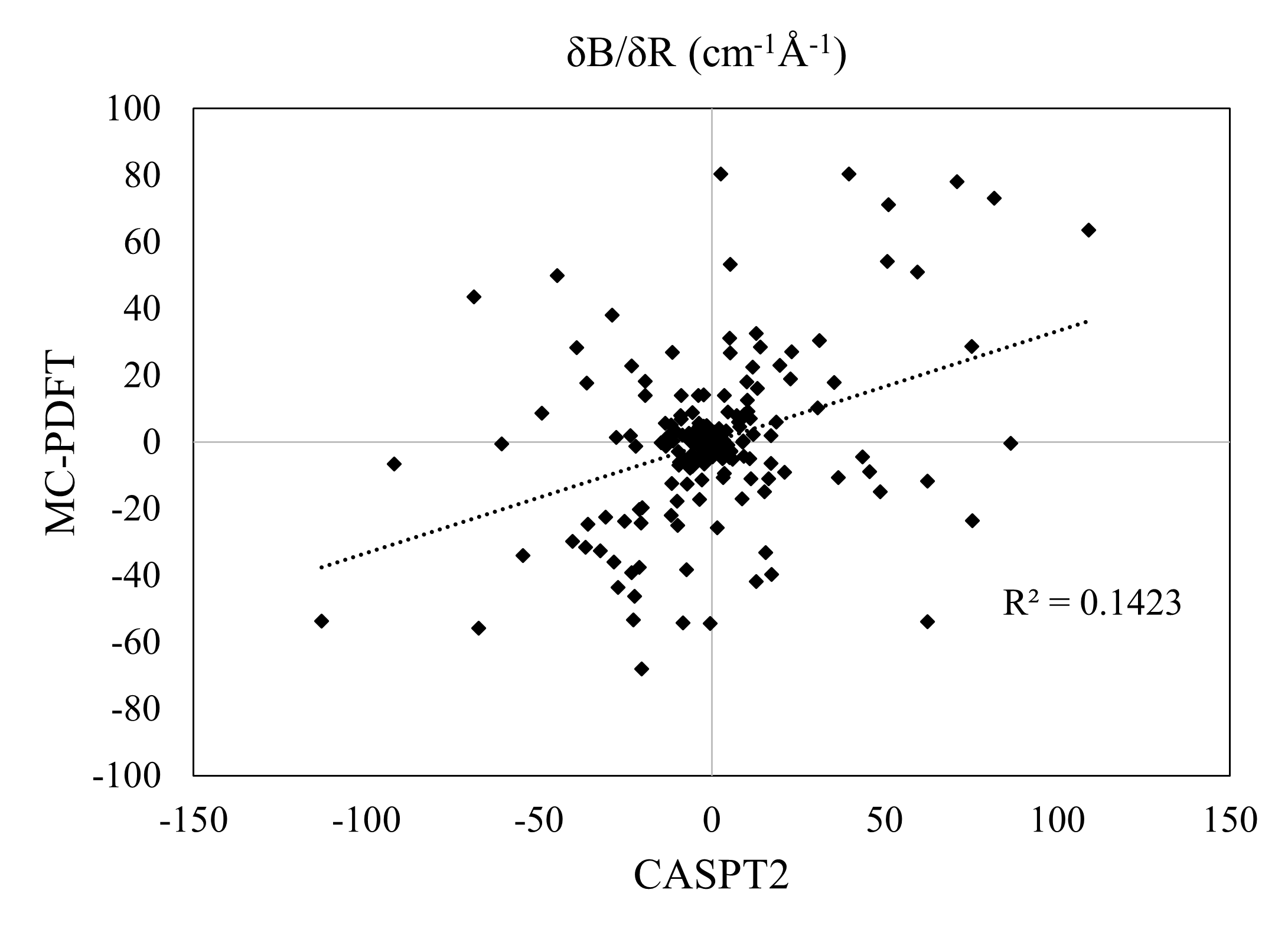}}
\centering
\caption{Parity plots comparing the numerical derivatives of the crystal field parameters computed at CASPT2 and MC-PDFT levels for compound \textbf{1}}
\label{fig: Parity plot mol_1 CASPT2 vs MC-PDFT}
\end{figure}

\begin{table}[htbp]
  \begin{center}
  \caption{\textbf{Total (Raman and Orbach) spin-phonon relaxation time (in s) for complex \textbf{1} at different temperatures (in K)}}
  \label{tau vs T mol_1}
  \begin{tabular}{c c c c}

    \hline
    T($K$) & CASSCF &
    CASPT2 & MC-PDFT \\
    \hline
    65 & 2.23E-08 & 5.99E-09 & 4.04E-09 \\
    40 & 9.14E-07 & 3.51E-07 & 2.76E-07 \\
    35 & 3.20E-06 & 1.30E-06 & 1.12E-06 \\
    30 & 1.32E-05 & 4.96E-06 & 4.89E-06 \\
    25 & 4.99E-05 & 1.52E-05 & 1.56E-05 \\
    20 & 1.51E-04 & 4.16E-05 & 4.12E-05 \\
    15 & 4.82E-04 & 1.38E-04 & 1.32E-04 \\
    10 & 2.42E-03 & 7.53E-04 & 7.06E-04 \\
    9 & 3.79E-03 & 1.20E-03 & 1.12E-03 \\
    8 & 6.43E-03 & 2.06E-03 & 1.92E-03 \\
    7 & 1.22E-02 & 3.92E-03 & 3.66E-03 \\
    6 & 2.69E-02 & 8.70E-03 & 8.11E-03 \\
    5 & 7.36E-02 & 2.39E-02 & 2.23E-02 \\
    \hline
    \end{tabular}
    \end{center}
 \end{table}%

\graphicspath{{Figures/}}
\begin{figure}[h!]
\centerline{\includegraphics[width=0.8\textwidth,,height=0.52\textwidth]{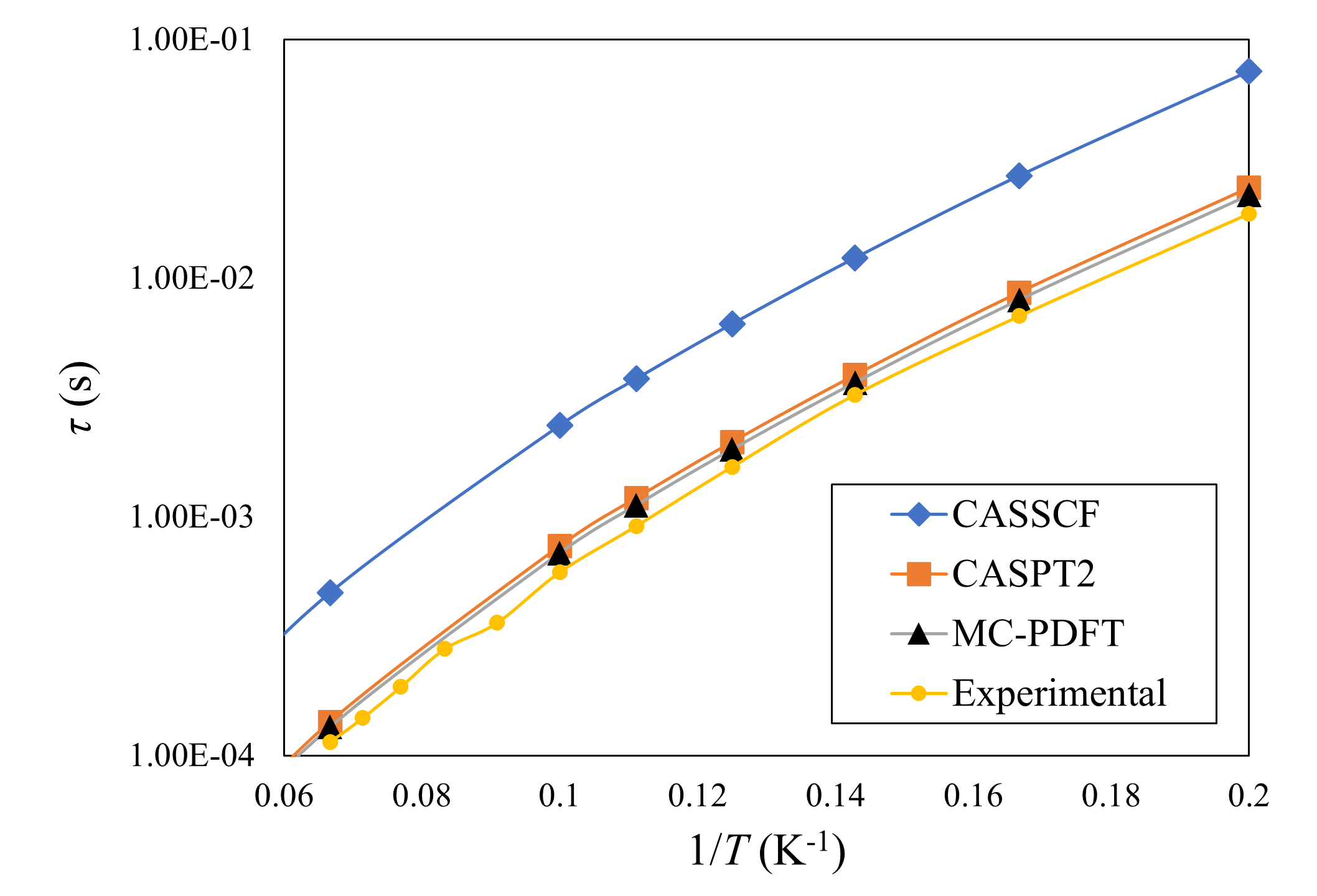}}
\centering
\caption{Enlarged portion of the total spin relaxation time as a function of 1/T for complex \textbf{1} obtained from different methods. Only the portion where experimental data is available is shown.}
\label{fig: spin-relaxation time mol_1 enlarged}
\end{figure}

\newpage

\section{Parity plots and total spin-relaxation time at different temperatures for [CoL\textsubscript{2}][(HNEt\textsubscript{3})\textsubscript{2}] (complex \textbf{2})}

\graphicspath{{Figures/}}
\begin{figure}[h!]
\centerline{\includegraphics[width=1.15\textwidth,,height=0.46\textwidth]{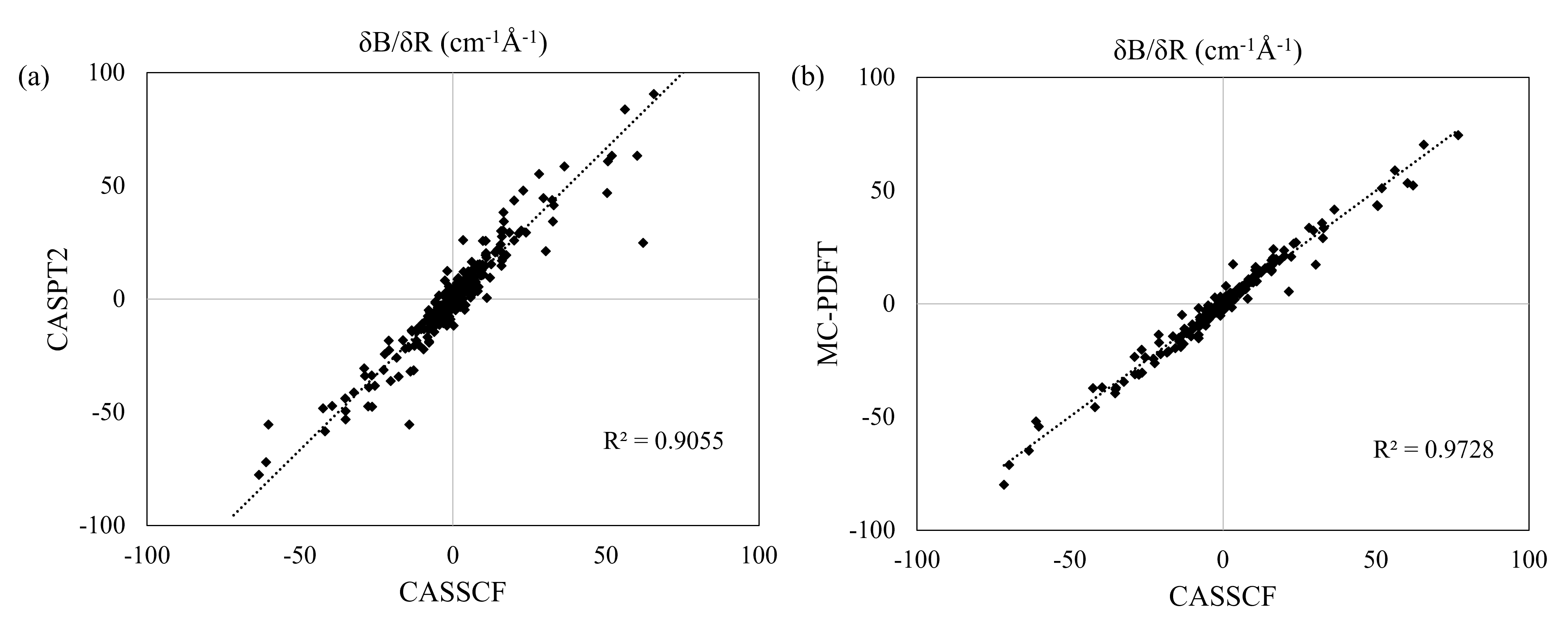}}
\centering
\caption{Parity plots comparing the numerical derivatives of the crystal field parameters computed at (a) CASSCF and CASPT2, and (b) CASSCF and MC-PDFT levels for compound \textbf{2}}
\label{fig: Parity plot mol_2}
\end{figure}

\begin{figure}[h!]
\centerline{\includegraphics[width=0.62\textwidth,,height=0.48\textwidth]{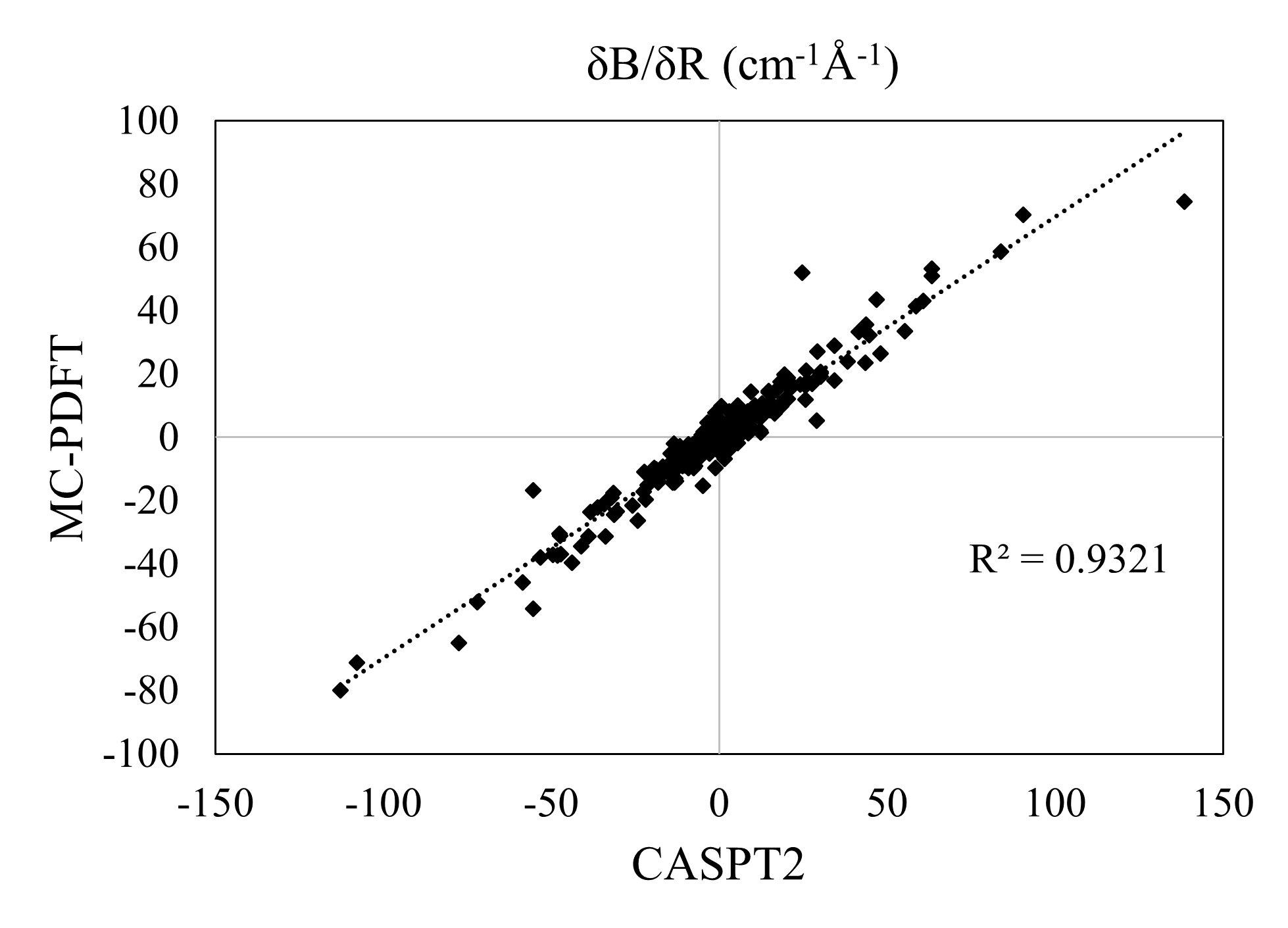}}
\centering
\caption{Parity plots comparing the numerical derivatives of the crystal field parameters computed at CASPT2 and MC-PDFT levels for compound \textbf{2}}
\label{fig: Parity plot mol_2 2}
\end{figure}

\begin{table}[htbp]
  \begin{center}
  \caption{\textbf{Total (Raman and Orbach) spin-phonon relaxation time (in s) for complex \textbf{2} at different temperatures (in K)}}
  \label{tau vs T mol_1}
  \begin{tabular}{c c c c}

    \hline
    T($K$) & CASSCF &
    CASPT2 & MC-PDFT \\
    \hline
    65 & 1.14E-08 & 5.39E-08 & 8.40E-09 \\
    40 & 1.98E-07 & 1.28E-06 & 1.16E-07 \\
    35 & 5.54E-07 & 3.74E-06 & 2.97E-07 \\
    30 & 2.07E-06 & 1.34E-05 & 1.00E-06 \\
    25 & 1.10E-05 & 5.44E-05 & 5.03E-06 \\
    20 & 6.91E-05 & 2.02E-04 & 4.17E-05 \\
    15 & 3.52E-04 & 7.76E-04 & 4.36E-04 \\
    10 & 2.69E-03 & 5.75E-03 & 3.42E-03 \\
    9 & 4.89E-03 & 1.04E-02 & 6.25E-03 \\
    8 & 9.95E-03 & 2.11E-02 & 1.28E-02 \\
    7 & 2.39E-02 & 5.04E-02 & 3.10E-02 \\
    6 & 7.24E-02 & 1.52E-01 & 9.48E-02 \\
    5 & 3.16E-01 & 6.58E-01 & 4.17E-01 \\
    \hline
    \end{tabular}
    \end{center}
 \end{table}%

\graphicspath{{Figures/}}
\begin{figure}[h!]
\centerline{\includegraphics[width=0.8\textwidth,,height=0.52\textwidth]{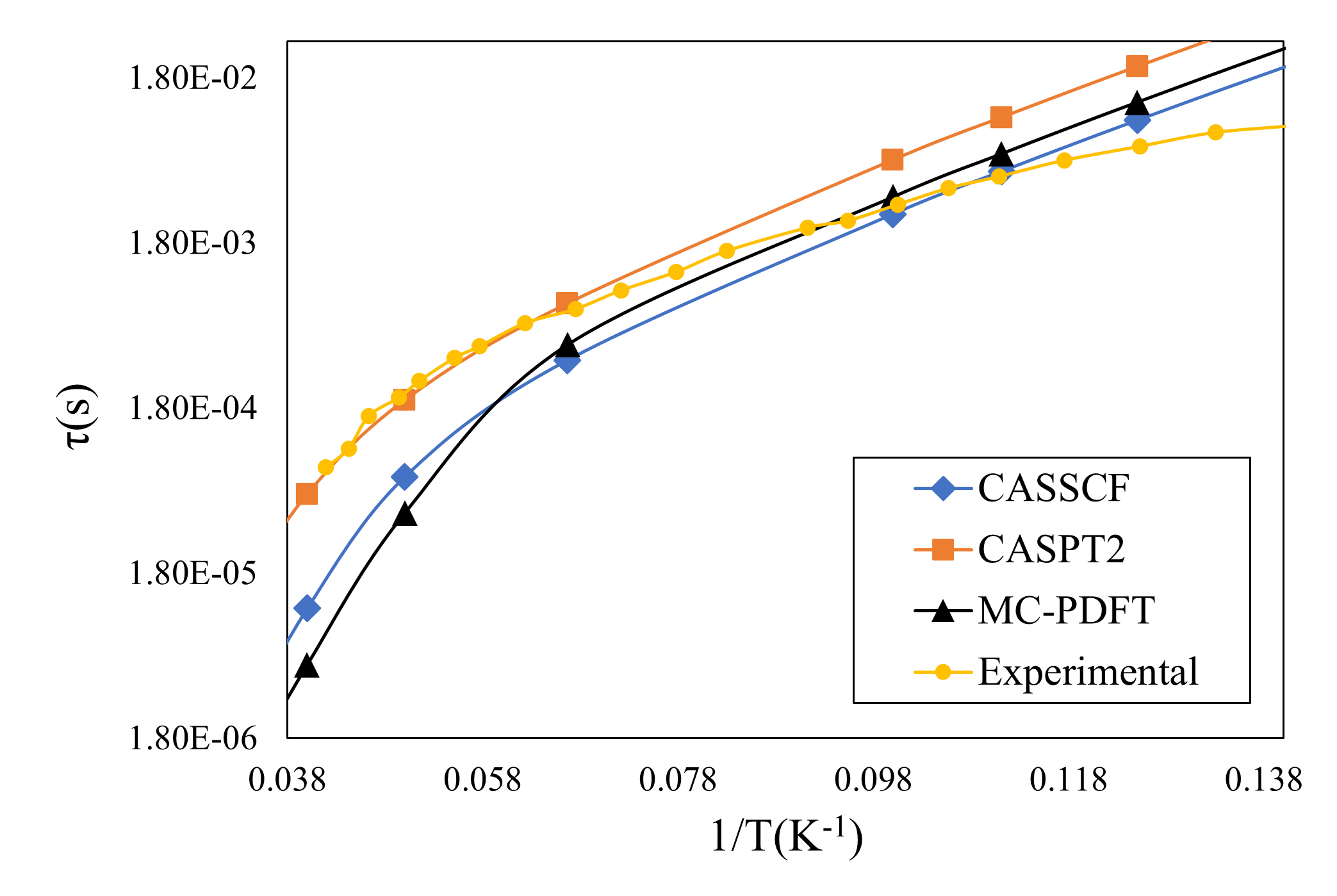}}
\centering
\caption{Enlarged portion of the total spin relaxation time as a function of 1/T for complex \textbf{2} obtained from different methods.}
\label{fig: spin-relaxation time mol_1 enlarged}
\end{figure}

\newpage

\section{Energies of Kramers doublets for [Dy(bbpen)Cl] (complex \textbf{3})}

\begin{table}
  \centering
  \caption{\textbf{Energies of the lowest Kramers doublets (in cm$^{-1}$) for complex \textbf{3} obtained from different electronic structure methods}}
  \label{tab: En KD mol_1}
  \begin{tabular}{>{\centering\arraybackslash}p{1cm} p{1.6cm} p{1.6cm} p{1.6cm} p{1.6cm} p{1.6cm} p{1.6cm} p{1.6cm} p{1.6cm}}
\hline
  \textbf{States} & CASSCF & CASSCF & CASPT2 & CASPT2 & MS-PT2 & XMS-PT2 & MC-PDFT & CMS-PDFT \\
   AS & (9e, 7o) & (9e, 14o) & (9e, 7o) & (9e, 14o) & (9e, 7o) & (9e, 7o) & (9e, 7o) & (9e, 7o) \\
  \hline
    KD$_0$ & 0 &
    0 & 0 & 0 & 0 & 0 & 0 & 0
     \\ 
    KD$_1$ & 383 & 401
 & 524 & 476 & 540 & 497 & 127 & 490 \\ 
    KD$_2$ & 610 & 638
 & 814 & 735 & 802 & 785 & 410 & 752 \\
    KD$_3$ & 700 & 728
 & 873 & 805 & 850 & 879 & 572 & 1042 \\
    KD$_4$ & 711 & 740
 & 909 & 827 & 858 & 898 & 880 & 1195 \\
    KD$_5$ & 743 & 774
 & 919 & 844 & 941 & 935 & 1001 & 1275 \\
    KD$_6$ & 781 & 816
 & 971 & 896 & 992 & 986 & 1202 & 1326 \\
    KD$_7$ & 827 & 861
 & 1020 & 945 & 1023 & 1038 & 1734 & 1513 \\
    \hline
    \end{tabular}
 \end{table}

\begin{table}[htbp]
  \begin{center}
  \caption{\textbf{Total (Raman and Orbach) spin-phonon relaxation time (in s) for complex \textbf{3} at different temperatures (in K)}}
  \label{tau vs T mol_1}
  \begin{tabular}{c c c}

    \hline
    T($K$) & CASSCF &
    CASPT2 \\
    \hline
    70 & 2.76E-07 & 9.83E-06 \\
    66 & 6.59E-07 & 3.04E-05 \\
    62 & 1.76E-06 & 1.08E-04 \\
    58 & 5.31E-06 & 4.37E-04 \\
    54 & 1.86E-05 & 1.97E-03 \\
    50 & 7.72E-05 & 7.83E-03 \\
    48 & 1.68E-04 & 1.31E-02 \\
    46 & 3.78E-04 & 1.92E-02 \\
    44 & 8.65E-04 & 2.54E-02 \\
    42 & 1.94E-03 & 3.21E-02 \\
    40 & 6.29E-03 & 4.04E-02 \\
    38 & 9.30E-03 & 5.07E-02 \\
    36 & 1.31E-02 & 6.44E-02 \\
    34 & 1.80E-02 & 8.31E-02 \\
    32 & 2.48E-02 & 1.09E-01 \\
    30 & 3.45E-02 & 1.46E-01 \\
    28 & 4.88E-02 & 1.99E-01 \\
    26 & 7.06E-02 & 2.77E-01 \\
    24 & 1.05E-01 & 3.96E-01 \\
    22 & 1.60E-01 & 5.79E-01 \\
    20 & 2.53E-01 & 8.74E-01 \\
    \hline
    \end{tabular}
    \end{center}
 \end{table}%

\section{Sample Inputs for Spin-phonon Relaxation Simulation using MolForge Software}

A sample input for the MolForge Spiral.x module used for the complex \textbf{1} is provided below. An elaborate description of the keywords can be found in MolForge manual (available at github.com/LunghiGroup/MolForge).

\&SPIN\_H \\
 \&DEF\_G  1 \#g-matrix of your system \\
2.0    0.0       0.0 \\
0.0      2.0   0.000 \\
0.0      0.0000    2.0 \\
 \&END \\
\&DEF\_O 1 2 \#Static  B coefficients \\
-2  -20.432430297018943 \\
-1  -75.542676238271810 \\
 0  -73.199339681285309 \\
 1  -40.405904383210917 \\
 2   12.417022878429290 \\
 \&END \\
EULER 1.0837654909188834 \\  0.60154069129323695   \\
2.6867033198157317 \\
\&END \\
\&SYSTEM \#Magnetic fields which has to be non-zero (0 0 0.3) for Raman relaxation and (0 0 0) for Orbach \\
 B 0.0000 0.0000 0.3000 \\
 \&DEF\_SPINS \#Not change \\
  S 1 1.5 -0.466867723 \\
 \&END \\
 \&CELL \#Not change \\
  A  100.00000  0.00000   0.00000 \\
  B  0.00000  100.00000   0.00000 \\
  C  0.00000    0.00000 100.00000 \\
  NREP 1 1 1 \\
  \&COORD \#Not change \\
   S 1 7.84704255057099    \\ 14.39220222767899  \\   
   5.00741750010644 \\
  \&END \\
 \&END \\
\&END \\
\&SPH\_H \\
 \&PHONDY \\
  TEMP 25 \# Temperature \\
  K\_MESH  1 1 1 \\
  SMEAR 25 \# Smearing can be varied more or less between ~5 and ~40. It should converge increasing from 0 to inf. \\
  SMEAR\_TYPE 1   \# 0=L 1=G \\
  FC2 FC2 \#File with phonons \\
  MAX\_ENER 3500 \#Energy window for the phonons. Has to be converged. \\
  MIN\_ENER 8 \\
 \&END \\
\&O\_BATH 1 2 \#File with B derivatives
FILENAME B\_final.txt \\
NORDER 1 \\
\&END \\
SECULAR \# Important! comment/uncomment if you want Orbach/Raman calculation. \\
PT2 \# Important! comment/uncomment if you want Orbach/Raman calculation. \\
\&END
\&HILBERT\_SPACE \#Not change \\
 fulldiag \\
 max\_ex -1 \\
 max\_corr -1 \\
 max\_dist 100000.0 \\
 dump\_freq 1 \\
 dump\_s T \\
 dump\_mi 1 1 \\
 dump\_rmat \\
\&END \\
\&DENSITY\_MATRIX \#Not change \\
TYPE FULLY\_POLARIZED \\
\&END DENSITY\_MATRIX \\
BUILD\_PROPAGATOR 1.00 1 \\
PROPAGATE 50000 \\